\documentclass[useAMS,usenatbib]{mn2e}
\usepackage{amsmath}
\usepackage{graphicx}
\newcommand{\ba}{\mbox{\boldmath{$\alpha$}}}
\newcommand{\bt}{\mbox{\boldmath{$\theta$}}}
\newcommand{\bb}{\mbox{\boldmath{$\beta$}}}
\newcommand{\f}{{\cal F}}
\newcommand{\g}{{\cal G}}
\newcommand{\bx}{{\mathbf x}}
\newcommand{\by}{{\mathbf y}}
\newcommand{\om}{\Omega_{\rm m}}
\newcommand{\ihat}{\hat{\mbox{\boldmath{$\imath$}}}}
\newcommand{\jhat}{\hat{\mbox{\boldmath{$\jmath$}}}}
\newcommand{\fmap}{F\!M_{\rm ap}}

\begin{document}

\label{firstpage}
\title[Masses and mass profiles from gravitational flexion]{A new tool to determine masses and mass profiles using gravitational flexion} \author[A. Leonard and L. J. King] {Adrienne
  Leonard\thanks{Email: leonard@ast.cam.ac.uk} and Lindsay
  J. King\\Institute of Astronomy and Kavli Institute for Cosmology, University of Cambridge, Madingley Road, Cambridge CB3 0HA}

\date{}
\maketitle

\pagerange{\pageref{firstpage}--\pageref{lastpage}} \pubyear{2009}

\begin{abstract}
In a recent publication, an aperture mass statistic for gravitational flexion was derived and shown to be effective, at least with simulated data, in detecting massive structures and substructures within clusters of galaxies. Further, it was suggested that the radius at which the flexion aperture mass signal falls to zero might allow for estimation of the mass or density profile of the structures detected. In this paper, we more fully explore this possibility, considering the behaviour both of the peak signal and the zero-signal contours for two mass models--the singular isothermal sphere and Navarro-Frenk-White profiles--under varying aperture size, filter shape and mass concentration parameter. We demonstrate the effectiveness of the flexion aperture mass statistic in discriminating between mass profiles and concentration parameters, and in providing an accurate estimate of the mass of the lens, to within a factor of $1.5$ or better. In addition, we compare the aperture mass method to a direct nonparametric reconstruction of the convergence from flexion measurements. We demonstrate that the aperture mass technique is much better able to constrain the shape of the central density profile, obtains much finer angular resolution in reconstructions, and does not suffer from ambiguity in the normalisation of the signal, in contrast to the direct method.
\end{abstract}


\section{Motivation}

There is an ongoing debate regarding the exact shape of galaxy- and cluster-scale dark matter haloes in the Universe. N-body simulations carried out under the assumption of the standard concordance cosmological model, $\Lambda$CDM, suggest that cold dark matter haloes are well fit by Navarro-Frenk-White (NFW) density profiles (Navarro, Frenk \& White, 1997), which are defined by a characteristic virial radius, $R_{200}$ (or, equivalently, the virial mass $M_{200}$), and a concentration parameter, $c$. However, gravitational lensing observations of field galaxies find that galaxy haloes consistently appear to show isothermal density profiles on the scales probed by strong lensing (Treu \& Koopmans, 2002; Rusin, Kochanek \& Keeton, 2003; Rusin and Kochanek, 2005; Koopmans et al., 2006; Gavazzi et al., 2007; Czoske et al., 2008; Dye et al., 2008; Tu et al., 2009), resulting from the interplay between baryons and dark matter. 
There is further debate concerning the density profiles of galaxy clusters (Carlberg et al., 1997; van der Marel et al., 2000; Athreya et al., 2002; Katgert, Biviano \& Mazure, 2004; Lin, Mohr \& Stanford, 2004; Hansen et al., 2005; {\L}okas et al., 2006; Rines \& Diaferio, 2006; Wojtak et al., 2007; Okabe et al., 2009); although some gravitational lensing studies favour an NFW model over an isothermal model, there are claims that NFW models give a poor fit to some clusters (e.g. Newman et al., 2009).

In addition, there has been much discussion regarding the relationship between halo mass and NFW concentration parameter. $\Lambda$CDM N-body simulations imply that the NFW profile is, in fact, a one-parameter (rather than two-parameter) model with the concentration parameter, $c$, being related directly to the virial mass $M_{200}$ as a power law with negative index (e.g. Navarro et al., 1997; Shaw et al., 2006). Gravitational lensing observations involving galaxy-galaxy lensing in the Sloan Digital Sky Survey, for example, seem to support this supposed mass-concentration relation (Mandelbaum, Seljak, \& Hirata, 2008). However, numerous gravitational lensing studies of clusters of galaxies, both in the strong and weak lensing regimes, have found significantly higher concentration parameters than would be expected from the mass-concentration relation (e.g. Abell 1689, for which a compilation of results can be found in Corless, King \& Clowe, 2009). 

Furthermore, in a number of gravitational lensing studies of clusters of galaxies, different sets of lensing measurements are seen to disagree, and different (NFW) parametric models of the cluster with widely varying masses and concentration parameters are seen by different authors to provide good fits to the data. For example, gravitational lensing studies of the galaxy cluster Abell 1689 (see, e.g., Corless et al. 2009; Peng et al. 2009; and references therein) have found best-fit models that span a factor of 3.5 in mass and 3 in best-fit NFW concentration parameter. Such disagreement hinders efforts to compare cluster data with the predictions from N-body simulations described above, and thus to constrain cosmological parameters. The technique presented in this paper is designed to assist in breaking some of these degeneracies, as even within a family of models such as the NFW model, the $\fmap$ signatures for profiles with different lens masses and concentration parameters are seen to be substantially different from one another.


Measurements of gravitational flexion, the second order gravitational lensing effects which give rise to skewness and arciness in galaxy images, have been shown by numerous authors to be adept at detecting galaxy- and galaxy group-size haloes both in the field and within clusters of galaxies (e.g. Okura, Umetsu \& Futamase, 2007,2008; Leonard et al., 2008; Leonard, King \& Wilkins, 2009), as well as providing an alternative method by which the mass distribution within a cluster of galaxies might be constrained. In addition, Lasky \& Fluke (2009) have found that the first flexion signal from an SIS profile differs from that of an NFW substantially at moderate separation between source and lens, whilst the second flexion signal shows strong variation when the NFW concentration parameter is varied. Flexion studies could therefore provide complementary constraints on halo profiles as well as their masses.

Recently, Leonard et al. (2009; hereafter LKW09) developed an aperture mass statistic for flexion, in direct analogy with that used for shear, and showed that it provided a robust method by which structures within clusters of galaxies, on a range of physical scales, can be identified to appreciable signal to noise. This technique is formally identical to the standard shear aperture mass methods used in weak lensing (see, for example, Schneider 1996, Schneider et al. 1998), but uses measurements of flexion rather than shear. This technique has the advantages that the noise properties of the aperture mass maps generated are very well understood, and that the filter functions used can be tuned to provide optimal signal-to-noise for a given lens profile. 

Moreover, the flexion aperture mass technique offers a robust method by which the mass distribution of a cluster of galaxies might be mapped out without the need for parametric modelling. This means that the technique does not rely on assumptions about the mass profiles of the structures responsible for the lensing signal, nor does it require \textit{a priori} knowledge of the locations of the mass concentrations responsible for the lensing signal. It can therefore be used to place independent constraints on the mass distribution of clusters of galaxies, without the invocation of any assumptions regarding the shape of the mass density profile of the structure.

In addition, LKW09 noted that the zero-signal contours expected to be found around mass peaks vary in radius under change of mass profile, lens mass, filter shape and aperture size. LKW09 suggest the properties of this set of contours might thus be used to provide insights into the mass and profile shape of the structures identified using this method, thus offering the potential to discriminate between competing mass models in cases where degeneracies arise. In this paper, we more fully investigate this claim. 

We focus on two properties of the flexion aperture mass signal--namely, the peak signal associated with a given structure and its zero-signal contour--and investigate their behaviour for SIS and NFW profiles under variation in virial mass, NFW concentration parameter, aperture radius and filter shape in order to quantify the discriminating power of this technique. We demonstrate that the flexion aperture mass signatures of these two mass profiles are very different under changes to the aperture and filter properties. As a result of this divergent behaviour, aperture mass filtering of flexion data using a variety of aperture parameters provides a convenient and straightforward method for estimating both the mass and profile shape of the structures detected without the need for parametric modelling. Moreover, we demonstrate on simulated data that the total mass can be constrained to within a factor of $\sim 1.5$ for both galaxy-group and cluster scale haloes, and the input profile shape recovered, for a signal to noise in the flexion aperture mass measurement as low as 1. Finally, we demonstrate that this method significantly outperforms a Fourier transform-based direct inversion technique (similar to that used in Okura et al.'s 2008 analysis of Abell 1689), yielding both higher resolution and much better constraints on the shape of the cluster mass profile.

This paper is structured as follows. In \S~\ref{sec:formalism}, we provide a brief review of the origin of the flexion signal, an overview of the flexion aperture mass statistic, and a description of the mass models considered. In \S~\ref{sec:radprof}, we consider the radial profile of 
the aperture mass  for each of the mass models and for a range of filters and aperture radii, and for varying values of the lens virial mass and concentration parameter (in the case of NFW lenses). The behaviour of the peak signal and the zero-signal contour under changes to the underlying mass model and changes to the aperture and filter properties is considered. In \S~\ref{sec:discrim}, we investigate the behaviour of a single mass profile under changes to the aperture and filter properties, and examine whether the use of several different combinations of aperture radius and filter shape in aperture mass reconstructions enables one to discriminate between different halo masses and density profiles. In \S~\ref{sec:nfw}, we apply aperture mass and direct reconstruction techniques to flexion data obtained by raytracing simulations through a cluster extracted from the Millennium simulation (see LKW09 and references therein) and compare the performance of the two methods. We conclude in \S~\ref{sec:summary} with a discussion of our results and their implications for future flexion studies. 

Throughout the text, we assume a standard $\Lambda$CDM cosmology with $\om=0.27$, $\Omega_\Lambda=0.73$, and $H_0=100h$\,km\,s$^{-1}$\,Mpc$^{-1}$.

\section{Basic Formalism}
\label{sec:formalism}
\subsection{Gravitational Flexion}
\label{subsec:flexion}

We begin by providing a very brief introduction to the flexion formalism. For a more complete discussion of the theory and description of the various techniques used to measure flexion, the reader is referred to Goldberg \& Bacon (2005), Bacon et al. (2006), Goldberg \& Leonard (2007), Leonard et al. (2008) and Okura et al. (2007, 2008). 

Gravitational flexion arises when the lens field varies significantly over the scale of the lensed image. In this case, the lens equation -- taken to be linear in weak lensing studies -- must be extended to second order:
\begin{equation}
\beta_i\simeq A_{ij}\theta_j+\frac{1}{2}D_{ijk}\theta_j\theta_k\ ,
\end{equation}
where $\bb$ is the coordinate in the source plane and $\bt$ is the coordinate in the lens plane. These coordinates are related by $\bb=\bt-\ba(\bt)$, where $\ba(\bt)$ is the deflection angle induced by the lens potential. {\bf A} is the magnification matrix, which is related to the convergence, $\kappa$, and the complex gravitational shear $\gamma=|\gamma|e^{2i\phi}$, and $D_{ijk}=\partial_kA_{ij}$ is related to the first flexion, $\f=|\f|e^{i\phi}=\partial\kappa$, and second flexion, $\g=|\g|e^{3i\phi}=\partial\gamma$, only. 

The first flexion signal gives rise to a skewness in the brightness distribution of the lensed galaxy image, and is a direct probe of the local gradient of the convergence. As such, it offers an ideal probe of cluster substructure, as it tends to be more sensitive to small-scale structures than to the large-scale cluster potential outside of the critical region of the cluster. 

The second flexion signal gives rise to a bending or arciness in the lensed galaxy image. Although we expect the second flexion signal to be larger for most mass profiles than the first flexion signal (see, e.g., Goldberg \& Bacon, 2005; Lasky \& Fluke, 2009), Goldberg \& Leonard (2007) and Leonard et al. (2008) have found second flexion to be significantly more difficult to measure. We therefore restrict our discussion for the remainder of this paper to measurements of first flexion. The convergence is related to the first flexion through:
\begin{equation}
\label{eq:kapx}
\kappa(\bx)=\frac{1}{2\pi}\Re\left[\int\ d^2\bx^\prime\ E_\f^\ast(\bx-\bx^\prime)\f(\bx^\prime)\right]+\kappa_0\ ,
\end{equation}
where
\begin{equation}
E_\f=\frac{1}{X^\ast},
\end{equation}
and $X=x_1+ix_2$. 

\subsection{Direct Reconstruction Techniques}
\label{subsec:direct}

To date, two groups have presented gravitational flexion studies of a cluster of galaxies with encouraging results. Leonard et al. (2008) presented the first measurements of flexion in a cluster of galaxies, and described a parametric modelling technique to reconstruct the distribution of mass within the galaxy cluster Abell 1689. Whilst this method showed good results in the periphery of the cluster, allowing mapping of the substructure content of the cluster, it was not able to constrain the central mass profile at all as a result of various masking techniques employed by the authors.

Okura et al. (2008) demonstrated a nonparametric flexion reconstruction technique on the same cluster, with good results being obtained using only 5 background sources per square arcminute and a positive detection in the central region of the cluster. 
Their method involves bin-averaging the flexion signal and reconstructing the convergence according to Equation \ref{eq:kapx}. This is performed most simply and efficiently in Fourier space, where the relationship is expressed as (Bacon et al., 2006):
\begin{equation}
\tilde{\kappa}=\frac{ik_1}{k_1^2+k_2^2}\tilde{\f_1}+\frac{ik_2}{k_1^2+k_2^2}\tilde{\f_2}\ ,
\end{equation}
where $\tilde{X}$ represents the Fourier transform of $X$. In taking the Fourier transform of $\tilde{\kappa}$ one recovers $\kappa-\kappa_0$, where $\kappa_0$ is a constant of integration usually set by requiring the convergence to go to zero at large distances from the cluster centre.

Such reconstructions may be problematic, however, as the resolution one can obtain in the output convergence map is strongly dependent on the background source density and distribution. Moreover, in the presence of significant substructure, bin-averaging over large pixels may have the effect of washing out the flexion signal. In addition, any masking schemes that result in a deficiency of sources in the central region of the cluster will result in the reconstructed convergence in this region being underestimated. It is for this reason that a parametric technique was chosen by Leonard et al. (2008) as a more appropriate method for reconstructing the mass distribution in Abell 1689 from flexion measurements.

\subsection{The Flexion Aperture Mass Statistic, $\fmap$}
\label{subsec:apmass}

Aperture measures have long been used in weak lensing, and the effect of aperture filtering is to provide a robust measure of the mass distribution within lens systems whilst simultaneously providing a straightforward method for determining the noise properties of the mass maps generated. In this way, aperture filtering offers an alternative nonparametric reconstruction method, which allows one to detect and map out mass concentrations within lens systems without requiring \textit{a priori} knowledge or assumptions regarding the shape of the mass profile of the lens.

The aperture mass statistic is defined as (Schneider 1996; LKW09):
\begin{equation}
\label{eq:apmass}
m(\bx_0)= \int d^2\bx\ \kappa(\bx+\bx_0)\ w(|\bx|).
\end{equation}
Using Equation \ref{eq:kapx} above, we can re-define this in terms of the measured flexion as 
\begin{eqnarray}
  \lefteqn{m(\bx_0)=}\nonumber\\
  & &\frac{1}{2\pi}\left(\Re \left[\int d^2\bx\ w(x)\int
      d^2\bx^\prime\ E_\f^\ast
      (\bx-\bx^\prime+\bx_0)\f(\bx^\prime)\right] \right.\nonumber\\
  & & \left.+\int d^2\bx\ w(x)\kappa_0\right).
\end{eqnarray}
This can be made
independent of the constant quantity $\kappa_0$ if we require that the
mass filter function $w(x)$ is compensated, i.e.
\begin{equation}
\label{eq:comp}
\int_0^\infty x\ w(x) dx=0.
\end{equation}

Making the transformation $\by=\bx^\prime-\bx_0$, and writing $d^2\bx
= x\ dx\ d\phi$, we obtain
\begin{eqnarray}
\label{eq:starthere}
\lefteqn{m(\bx_0)=}\nonumber\\
& &\Re\left[\int d^2\by\f(\by+\bx_0)\int_0^\infty xw(x)dx
  \int_0^{2\pi}d\phi\ E_\f^\ast(X-Y)\right],\nonumber\\
\end{eqnarray}
Applying the residue theorem to the integral over $\phi$ yields (see Appendix B of LKW09 for a full derivation)
\begin{eqnarray}
m(\bx_0)=-\Re\left[\int d^2\by\  \f(\by-\bx_0)\frac{1}{Y}\int_0^y x\ w(x)\ dx
\right].
\end{eqnarray}

The $\fmap$-statistic can now be redefined in terms of the ``E-mode'' (radially aligned) flexion as
follows:
\begin{equation}
\label{eq:mflex}
m(\bx_0)=\int d^2\by\ \f_E(\by;\bx_0)\ Q_\f(y),
\end{equation}
where
\begin{eqnarray}
\f_E&=&\Re\left[\f e^{-i\phi}\right]\nonumber\\
&=& \Re\left[\f \frac{|Y|}{Y}\right],
\end{eqnarray}
and
\begin{equation}
\label{eq:qwflex}
Q_\f(y)=-\frac{1}{y}\int_0^y\ x\ w(x)\ dx.
\end{equation}
We note that the E-mode flexion can equivalently be expressed as
\begin{equation}
\f_E=\f\cdot\hat{\by}\ ,
\end{equation}
where here we have expressed the first flexion as a vector, rather than in complex notation, and $\hat{\by}\equiv \by/|\by|$.

As discussed in LKW09, any set of filters that are continuous and satisfy Equations \ref{eq:comp} and \ref{eq:qwflex} can be used in an $\fmap$ analysis. Ideally, one would choose a flexion filter function, $Q_\f(y)$, that traces the expected flexion signal in order to optimise the signal to noise in the $\fmap$ maps generated. However, for the purposes of this paper, we assume that the mass profile is not known \textit{a priori}. We choose the family of polynomial filters in LKW09, which are given by
\begin{equation}
\label{eq:qf}
Q_\f(x)=
-\frac{2(2+l)}{\pi^{\frac{3}{2}}}\frac{\Gamma\left(\frac{7}{2}+l\right)}{\Gamma(3+l)}x\left(1-\frac{x^2}{R^2}\right)^{1+l},
\end{equation}
where $R$ is the radius of the aperture being used. This family of filter functions has been shown by LKW09 to offer appreciable signal to noise for a range of different mass profiles, and is robust in the structures detected under variation of polynomial order $l$ and aperture radius $R$. 

\subsection{Mass Models}

We consider two of the most commonly-used models for gravitational lenses: the singular isothermal sphere (SIS) and Navarro-Frenk-White (NFW) profiles. Lasky and Fluke (2009) have presented an extensive description of the convergence, shear and flexion signals from both these models, as well as a S\'{e}rsic profile, and we use their results -- transformed into angular coordinates -- in this paper. In addition, we assume our lenses to be circularly symmetric. Hawken and Bridle (2009) have presented a description of flexion from elliptical lenses; however it is sufficient for the purposes of this paper, and indeed makes the analysis somewhat simpler, to consider only lenses with circular symmetry.

In order to compare these models, it is first helpful to define the virial radius, $R_{200}$, and the virial mass, $M_{200}$. The virial radius is defined to be the radius within which the average density $\overline{\rho}(R_{200}) = 200\rho_c$, where
\begin{equation}
\rho_c=\frac{3H^2}{8\pi G}
\end{equation}
is the critical density of the Universe, and
\begin{eqnarray}
\overline{\rho}(r)&=&\frac{3}{4\pi r^3}\int \rho(x) d^3x\nonumber\\
&=&\frac{3}{r^3}\int_0^r\ \rho(x)\ x^2\ dx
\end{eqnarray}
for a spherically symmetric lens. The virial mass is then given by
\begin{equation}
\label{eq:mvir}
M_{200}=\frac{800\pi}{3}\rho_c R_{200}^3.
\end{equation}

We also remind the reader that the convergence, $\kappa$, of a lens is related to its 2-dimensional projected surface mass density, $\Sigma$, by
\begin{equation}
\kappa(\bt)=\frac{\Sigma(\bt)}{\Sigma_c}\ ,
\end{equation}
where the angular coordinate, $\bt$, is related to the physical projected position vector, $\mbox{\boldmath{$\xi$}}$, and the observer-lens separation, $D_{\rm d}$, by $\bt=\mbox{\boldmath{$\xi$}}/D_{\rm d}$, and 
\begin{equation}
\Sigma_c=\frac{c^2}{4\pi G}\frac{D_{\rm s}}{D_{\rm ds}D_{\rm d}},
\end{equation}
where $D_{\rm s}$ is the observer-source separation and $D_{\rm ds}$ is the lens-source separation\footnote{All distances used are angular-diameter distances, unless otherwise specified.}.

\subsubsection{Singular Isothermal Sphere Profile}

The SIS model has a 3-D density profile given by
\begin{equation}
\rho(r)=\frac{\sigma^2}{2\pi Gr^2}\ ,
\end{equation}
where $\sigma$ is the velocity dispersion of the lens. The mean density within radius $r$ is therefore given by
\begin{equation}
\overline{\rho}(r)=\frac{3\sigma^2}{2\pi G r^2}\ .
\end{equation}
Setting this equal to $200\rho_c$, we find 
\begin{equation}
R_{200}=\frac{\sigma}{\sqrt{50}H}\ ,
\end{equation}
which gives 
\begin{equation}
M_{200}= \frac{2\sigma^3}{\sqrt{50}GH}\ .
\end{equation}
Projecting onto the lens plane, the convergence is given by
\begin{equation}
\kappa(\bt)=\frac{\theta_E}{2|\bt|},
\end{equation}
where $\theta_E$ is the Einstein radius of the lens, given by
\begin{equation}
\theta_E=4\pi\left(\frac{\sigma}{v_c}\right)^2\frac{D_{\rm ds}}{D_{\rm s}},
\end{equation}
$v_c$ is the velocity of light, and the first flexion is given by
\begin{equation}
{\cal F}(\bt)=-\frac{\theta_E}{2|\bt|^2}\hat{\bt}\ .
\end{equation}

\subsubsection{Navarro-Frenk-White Profile}

The NFW profile has a 3-D density defined by 
\begin{equation}
\rho(r)=\frac{\delta_c\rho_c}{(r/r_s)(1+r/r_s)^2}\ ,
\end{equation}
where
\begin{equation}
\delta_c=\frac{200}{3}\frac{c^3}{\ln(1+c)-c/(1+c)},
\end{equation}
and $c$ is the concentration parameter, defined as $c=R_{200}/r_s$. For a given choice of virial mass, the virial radius $R_{200}$ is computed from the definition in Equation \ref{eq:mvir}.

Before defining the convergence, it is useful to introduce some shorthand notation. First, we define $x\equiv |\mbox{\boldmath{$\xi$}}|/r_s=|\bt|/\theta_s$, where $|\bt|=|\mbox{\boldmath{$\xi$}}|/D_{\rm d}$. Second, we define a normalisation factor:
\begin{equation}
\kappa_c=\frac{2\rho_c\delta_cr_s}{\Sigma_c}\ .
\end{equation}
In this notation, the convergence is given by
\begin{equation}
\kappa(x)=\frac{\kappa_c}{\left(x^2-1\right)}\left[1-\Xi(x)\right],
\end{equation}
where
\begin{equation}
\Xi(x)=\begin{cases}
\frac{2}{\sqrt{1-x^2}} \arctan $h$\left(\sqrt{\frac{1-x}{1+x}}\right) & {\scriptstyle x<1}\\ \\
\frac{2}{\sqrt{x^2-1}}\arctan\left(\sqrt{\frac{x-1}{x+1}}\right) & {\scriptstyle x>1}
\end{cases},
\end{equation} 
and the first flexion is given by
\begin{equation}
{\cal F}(x)=-\frac{\kappa_c}{\theta_sx(x^2-1)^2}\left(2x^2+1-3x^2\Xi(x)\right)\hat{\bt}\ .
\end{equation}

\section{The $\fmap$ Radial Profile}
\label{sec:radprof}

We now consider the expected behaviour of the 
$\fmap$ signal for a given mass model, aperture size and filter shape. For convenience, we shall consider only apertures that lie along the $x$-axis, i.e. the position vector from the centre of the lens to the centre of the aperture is given by ${\bf x_0}=x_0 \ihat$. This
is a valid simplification because the lenses under consideration all have circular symmetry, hence the radial profile of the $\fmap$ signal along any given axis will provide a complete description of the expected signal. The displacement vector of a point within the aperture with respect to the
centre of the aperture is denoted by ${\bf y}=y\cos\phi\ihat+y\sin\phi\jhat$. Thus, the separation between the centre of the lens and a given point within the aperture may be expressed as $\bt={\bf  x_0}+{\bf y}=(x_0+y\cos\phi)\ihat+y\sin\phi\jhat$.

The flexion vector always points along the direction of $\bt$, i.e.:
\begin{equation}
{\cal F}(\bt)=|{\cal F}|\frac{\bt}{|\bt|}\ ,
\end{equation}
and, by definition,
\begin{eqnarray}
{\cal F}_E({\bf y};{\bf x_0})&=&\frac{{\cal F}\cdot {\bf
    y}}{y}\nonumber\\ 
 &=&\frac{|{\cal
    F}|}{\sqrt{x_0^2+y^2+2x_0y\cos\phi}}\left[x_0\cos\phi+y\right]\ .
\end{eqnarray}
It is important to note that $|{\cal F}|$ will, itself, generally be
dependent on $x_0$, $y$, and $\phi$, as it is a function of $\bt$. Thus, rewriting Equation \ref{eq:mflex}, $\fmap$ is given by
\begin{eqnarray}
m({\bf
  x_0})&=&\frac{2(2+l)}{\pi^\frac{3}{2}}\frac{\Gamma\left(\frac{7}{2}+l\right)}{\Gamma(3+l)}\int_0^R\ y^2\left(1-\frac{y^2}{R^2}\right)^{1+l}\ dy\nonumber\\\nonumber\\ & &\int_0^{2\pi}\ \frac{|{\cal
    F}|(x_0,y,\phi)[x_0\cos\phi+y]}{\sqrt{x_0^2+y^2+2x_0y\cos\phi}}\ d\phi\ .
\end{eqnarray}

As can be seen above, in general 
the flexion aperture mass is rather difficult to evaluate analytically, even for the simple case of a singular isothermal lens (unless the aperture is centred on the lens; this special case is discussed below). Therefore, in order to characterise its behaviour under different mass models, aperture scales and filter shapes, it is necessary to evaluate the signal numerically. 

To do this, for each combination of virial mass, mass model, aperture size and filter shape, we consider 200 apertures evenly spaced in the range $x_0=[0'',200'']$. At each aperture location, we use a uniform random number generator to generate 1000 $\by$ positions with $-R\le y_i\le R$, and select those that lie within the aperture (i.e. those with $|\by|\le R$), so that the signal is evaluated on a uniform distribution of points within the aperture. At each point, the E-mode flexion with respect to the centre of the aperture is computed, and the $\fmap$ signal at $x_0$ evaluated according to
\begin{equation}
m(x_0)=\frac{1}{n}\sum_i\ \f_{E,i}\ Q(y_i),
\end{equation}
where $n$ is the number density of points within the aperture. 

\begin{figure}
\center
\includegraphics[width=0.45\textwidth]{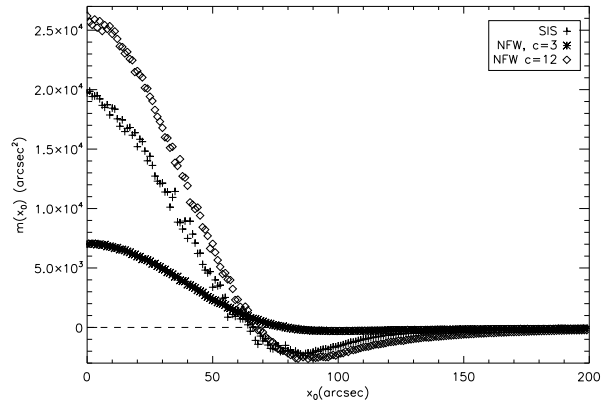}
\caption[Expected Aperture Mass Profile for a Cluster-Mass Lens]{The expected $\fmap$ signal, evaluated numerically, for a lens of $M_{200}=10^{15}h^{-1}M_\odot$ using a polynomial filter with $l=3$ and $R=120''$. The signal is plotted for an NFW profile with concentration parameter $c=3$ and $c=12$ and for an SIS profile. The dotted horizontal line shows $m(x_0)$=0.}
\label{fg:apmass_ex}
\end{figure}

As only a finite number of points are being evaluated, this measure can be rather noisy. To reduce the noise, for each value of $x_0$, the signal is estimated for 200 different sets of randomly-sampled data points, and the median value taken as an estimate of the signal. The resulting $\fmap$ signal is plotted in Figure \ref{fg:apmass_ex} for a sample of cluster mass lenses with $M_{200}=10^{15}h^{-1}M_\odot$, $l=3$ and $R=120''$. 

It is immediately apparent that the $\fmap$ profiles differ greatly both in amplitude and in effective width. Therefore, comparison of these two features should provide information about the underlying lens mass and density profile. We take the radius at which the signal goes to zero as a proxy for the effective width of the $\fmap$ profile in each case. Other contours, such as the width at half the maximum signal or the radius of the minimum signal, could be used; however, we selected the zero-signal contour as both easily identifiable and convenient to fit.

The profiles shown in the figure are not well-fit by a single analytic function. It is possible to fit these curves with a Chebyshev series of high order ($\sim 20$; T. Nguyen, \textit{priv. comm.}). However, the signal is equally well-fit (and, in some cases, better fit) by a piecewise continuous polynomial function, with a different polynomial behaviour seen for $x_0<x_{0,min}$ and for $x_0>x_{0,min}$, where $x_{0,min}$ is the value of $x_0$ at which the minimum signal is seen. We find that the data in the region $x_0<x_{0,min}$ are generally well fit by a polynomial of 
$4^{th}$ order in $x_0$. The zero-signal radius for an $\fmap$ profile is therefore determined by finding the first real, positive root of the best-fit 4$^{th}$ order polynomial.

In order to assess the behaviour of these two features of the $\fmap$ signal for different masses, mass profiles, aperture sizes and filter shapes, we carry out this numerical evaluation of the $\fmap$ signal for $10^{11}h^{-1}\,M_\odot \le M_{200} \le 10^{15}h^{-1}\,M_\odot$, $R=[45'',60'',75'',90'',105'',120'']$, and $l=[3,4,5,6,7,8,9,10]$. For each mass, we evaluate the signal for an SIS model, and for NFW models with concentration parameter $c=[3,4,5,6,7,8,9,10,11,12]$. Note that throughout this paper, all angular scales are measured in arc seconds; therefore $\fmap$ carries units of arcsec$^{2}$. 

In addition, for all the simulations carried out here, we assume a lens redshift of $z_{\rm d}=0.2$ and a source redshift of $z_{\rm s}=1.0$, typical of lens system geometries encountered in practice. These distances come into the normalisation of the flexion signals in different ways for the two mass profiles considered here. The singular isothermal sphere model has a flexion measurement related to the source and lens separations through:
\begin{equation}
\f_{\rm SIS}\propto \theta_E\propto\frac{D_{\rm ds}}{D_{\rm s}},
\end{equation}
whilst the redshift dependence of the NFW flexion signal is given by:
\begin{equation}
\f_{\rm NFW}\propto \frac{\rho_c r_s}{\Sigma_c}\propto H^2(z_{\rm d})\frac{D_{\rm ds}D_{\rm d}^2}{D_{\rm s}}\ ,
\end{equation} 
where $H(z_{\rm d})=H_0\sqrt{\Omega_{\rm M}(1+z_{\rm d})^3+\Omega_\Lambda}$. These are simple scaling relations, however, and it is straightforward to compare measurements with different source and lens redshifts by appropriately scaling the $\fmap$ signal.

\subsection{SIS model}
We now consider the behaviour of the peak $\fmap$ signal and the zero-signal radius for the SIS mass profile under variation of the aperture size and filter polynomial order. The peak signal, by definition located at $\bx_0=0$, can be evaluated analytically for the SIS model. For this model, at ${\bf x_0}=0$, we find
\begin{eqnarray}
{\cal F}_E&=&|{\cal F}|(y)\nonumber\\
&=& -\frac{\theta_E}{2y^2}\ .
\end{eqnarray}
\noindent $\fmap$ therefore becomes:
\begin{eqnarray}
m(0)&=&\theta_E \frac{2(2+l)}{\sqrt{\pi}}\frac{\Gamma\left(\frac{7}{2}+l\right)}{\Gamma(3+l)}\int_0^R\ \left(1-\frac{y^2}{R^2}\right)^{1+l}\ dy\nonumber\\
&=&\theta_E \frac{2(2+l)}{\sqrt{\pi}}\frac{\Gamma\left(\frac{7}{2}+l\right)}{\Gamma(3+l)}\left[\sqrt{\pi}R\frac{\Gamma(2+l)}{2\Gamma\left(\frac{5}{2}+l\right)}\right]\ .
\end{eqnarray}
Noting that
\begin{equation}
\Gamma(n+1)=n\Gamma(n)\ ,
\end{equation}
if $n$ is a positive integer or half-integer, we find that
\begin{equation}
m(0)=\theta_E\left(\frac{5}{2}+l\right)R\ .
\end{equation}
Expressed in terms of the virial mass of the lens, the expected peak $\fmap$ signal is given by:
\begin{equation}
\label{eq:mp_sis}
m_{peak}\equiv m(0)=4\pi\left(\frac{\sqrt{50}GHM_{200}}{2v_c^3}\right)^{\frac{2}{3}}\frac{D_{\rm ds}}{D_{\rm s}}\left(\frac{5}{2}+l\right)R\ .
\end{equation}

We now consider the point at which the $\fmap$ signal goes to
zero. This manifests in flexion aperture mass reconstructions as a fairly
well-defined zero-point contour (LKW09). It is not particularly
straightforward to solve for this zero-signal radius analytically; therefore this radius is evaluated numerically by fitting the data in the region $x_0<x_{0,min}$ with a 4$^{th}$ order polynomial in $x_0$, and subsequently finding the first real, positive root of this function.

The zero-signal contour for the SIS model appears to have no
dependence on mass. This is to be expected because the aperture mass
$m(\bx_0)\propto\theta_E\propto M_{200}^{\frac{2}{3}}$ for all values
of $\bx_0$. Moreover, the SIS model has no mass-dependent scale
radius. Therefore, in setting $m(\bx_0)=0$, the dependence
on $\theta_E$ is removed, and one would expect the zero-point radius to be related only
to the polynomial order $l$ of the filter and the aperture radius $R$. The data imply that the zero-point radius is well-fit by
\begin{equation}
R_0=\frac{R}{(2l)^{\frac{1}{3}}}
\end{equation}
for all masses.

\begin{figure}
\center
\includegraphics[width=0.45\textwidth]{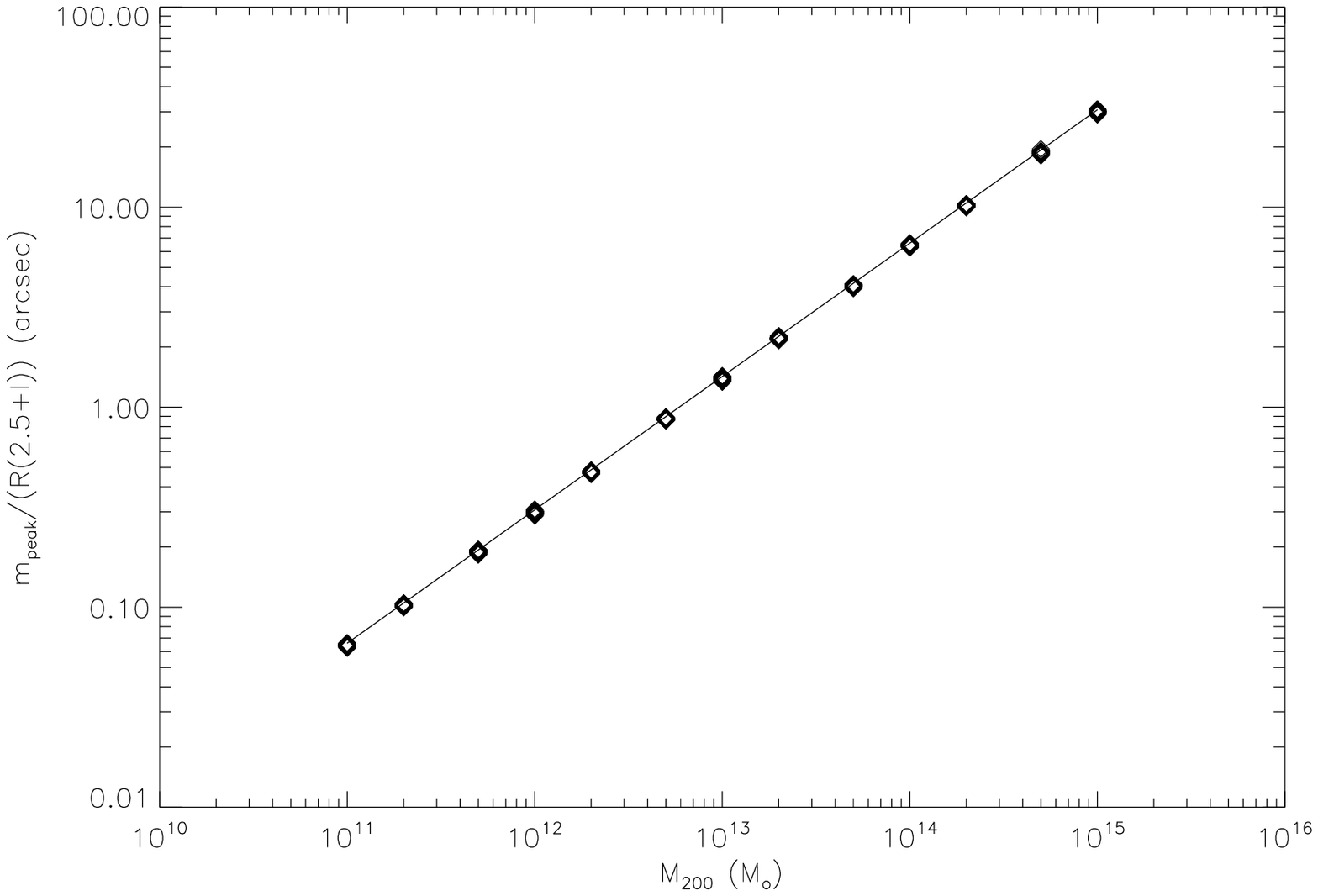}
\includegraphics[width=0.45\textwidth]{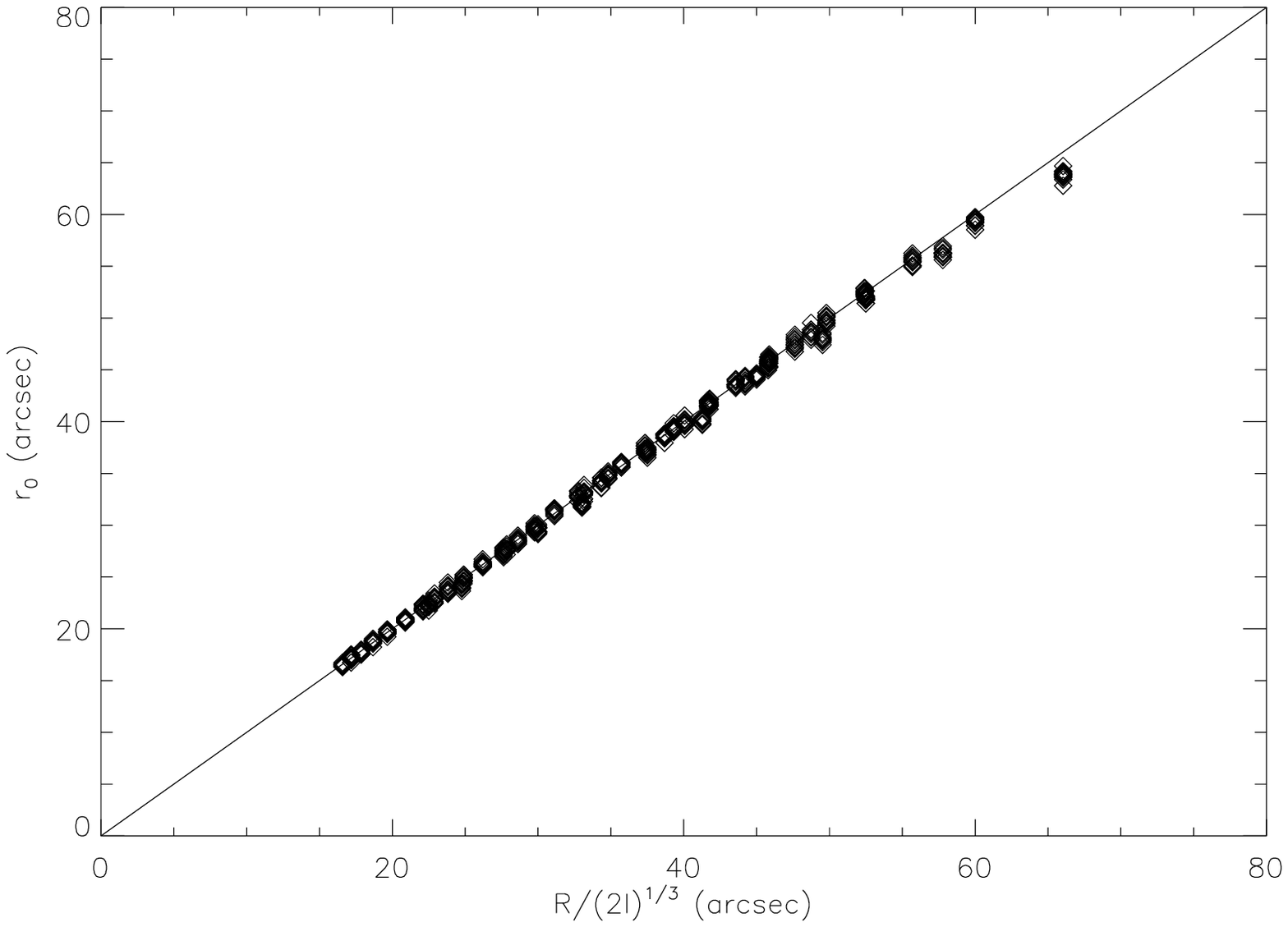}
\caption[Flexion $\fmap$ Signal for a Singular Isothermal Lens]{\textit{Top Panel:} The expected peak signal, normalised by aperture size $R$ and filter polynomial order $l$, as a function of $M_{200}$ for an SIS lens. The solid line shows the predicted signal from Equation \ref{eq:mp_sis}. \textit{Bottom Panel:} $R_0$ vs $R/(2l)^{1/3}$ for an SIS lens. The solid line represents a 1:1 mapping. Note that for each mass bin plotted, 48 different combinations of $l$ and $R$ were used. Thus, the plot above represents 624 data points in total.\label{fg:SIS}}
\end{figure}

Figure \ref{fg:SIS} shows a logarithmic plot of $M_{200}$ vs $m_{peak}/R\left(\frac{5}{2}+l\right)$ for the entire numerical data set, consisting of 48 different combinations of $l$ and $R$ in each of 13 mass bins. The solid line shows the predicted signal from Equation \ref{eq:mp_sis}. The correlation seen here provides an excellent test of the numerical methods used to evaluate the $\fmap$ signal. The figure also shows the behaviour of $R_0$ vs $R/(2l)^{1/3}$ for the numerical data set.

\subsection{NFW Model}

The NFW profile is rather more complicated, making a fully analytic calculation of the peak $\fmap$ signal for filters with polynomial order $l>0$, or for a number of different values of $l$, $R$, and $\theta_s$ simultaneously, impractical and not particularly instructive. Therefore, estimating the signal numerically, and fitting the resulting data with a somewhat simpler function, seems a more appropriate approach. Indeed, this is necessary for the calculation of $R_0$, which cannot be computed analytically even for a simple lens profile such as the SIS. The peak signal and zero-signal radius are both found to depend in a non-trivial way on the virial mass $M_{200}$, concentration parameter ($c$), aperture radius $R$ and filter polynomial order $l$. 

\begin{figure}
\center
\includegraphics[width=0.45\textwidth]{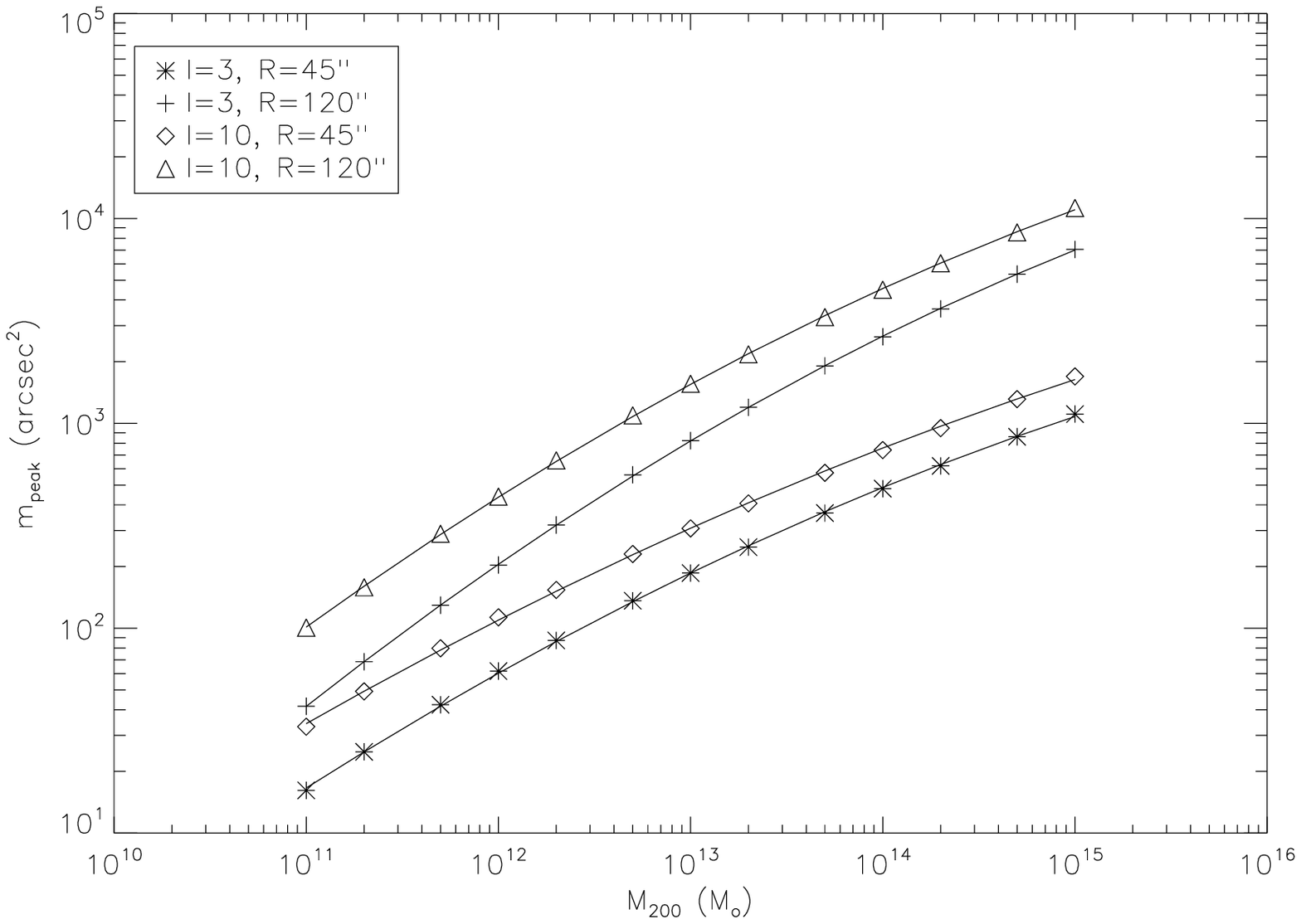}
\includegraphics[width=0.45\textwidth]{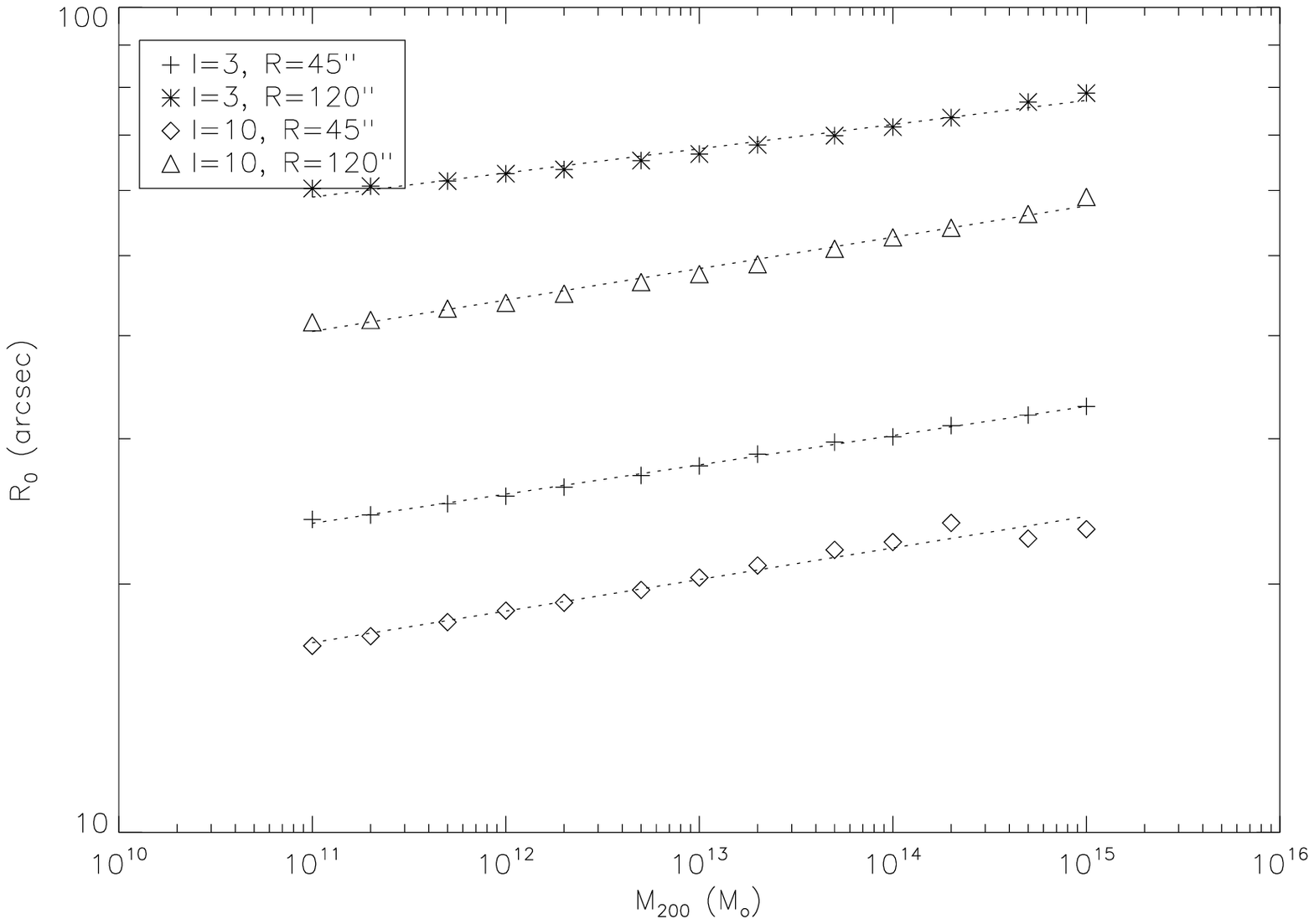}
\caption[Peak Flexion Aperture Mass Signal and Zero-signal Radius for an NFW Lens with $c=3$]{\textit{Top Panel:} The peak $\fmap$ signal as a function of mass for an NFW lens with concentration parameter $c=3$. The best fit curves of the form $\log_{10}(m_{peak})=a+b\log_{10}(M_{200})+d(\log_{10}(M_{200}))^2$ are shown as solid lines. \textit{Bottom Panel:} The zero-signal contour as a function of virial mass for an NFW lens with $c=3$. The power law fits, $R_0=AM_{200}^n$, are shown as dotted lines.\label{fg:nfw3m}}
\end{figure}

Figure \ref{fg:nfw3m} shows the behaviour of the peak signal and zero-signal radius as a function of $M_{200}$ at fixed $l$ and $R$ for an NFW profile with a concentration parameter of $c=3$. We find that the peak signal data are well fit by a polynomial of the form $\log_{10}(m_{peak})=a+b\log_{10}(M_{200})+d(\log_{10}(M_{200}))^2$, whilst the zero-signal radius $R_0$ appears to follow a power law of the form $R_0=AM_{200}^n$. The figure also shows the best fit curves, the parameters of which are given in Table \ref{tab:fitparams}.

\begin{table}
\centering
\begin{tabular}{c c c c c c c}
\hline
$l$ & $R('')$ & $a$ & $b$ & $d$ & $\log_{10}(A)$ & $n$\\
\hline
\ 3 & \ 45 & \ \ -9.6455 & 1.3799 & -0.0356 & 0.9846 & 0.0355\\
\ 3 & 120 & -12.0236 & 1.7413 & -0.0456 & 1.4476 & 0.0293\\
10 & \ 45 & \ \ -7.7901 & 1.1614 & -0.0286 & 0.8093 & 0.0383\\
10 & 120 & -10.4559 & 1.5902 & -0.0416 & 1.1886 & 0.0381\\
\hline
\end{tabular}
\caption[Peak Signal and Zero-signal Radius vs $M_{200}$ Power Law Fit Coefficients for an NFW Lens]{Table of coefficients for the best fit curves for the data shown in Figure \ref{fg:nfw3m}: the peak $\fmap$ signal and zero-signal contours vs $M_{200}$ for an NFW model with a concentration parameter of $c=3$. The fits are of the form  $\log_{10}(m_{peak})=a+b\log_{10}(M_{200})+d(\log_{10}(M_{200}))^2$ and $R_0=AM_{200}^n$.\label{tab:fitparams}}
\end{table}

\begin{figure}
\center
\includegraphics[width=0.45\textwidth]{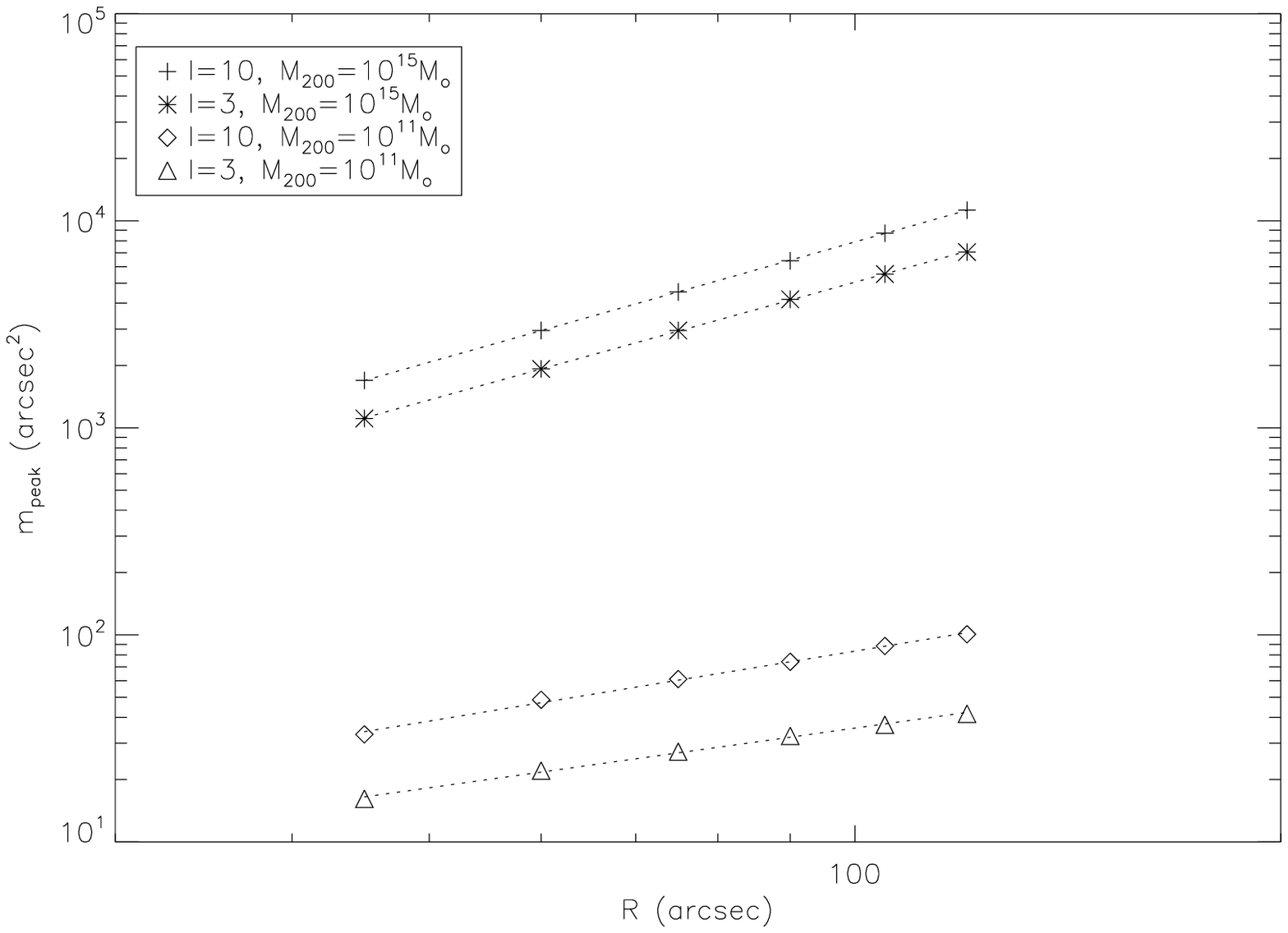}
\includegraphics[width=0.45\textwidth]{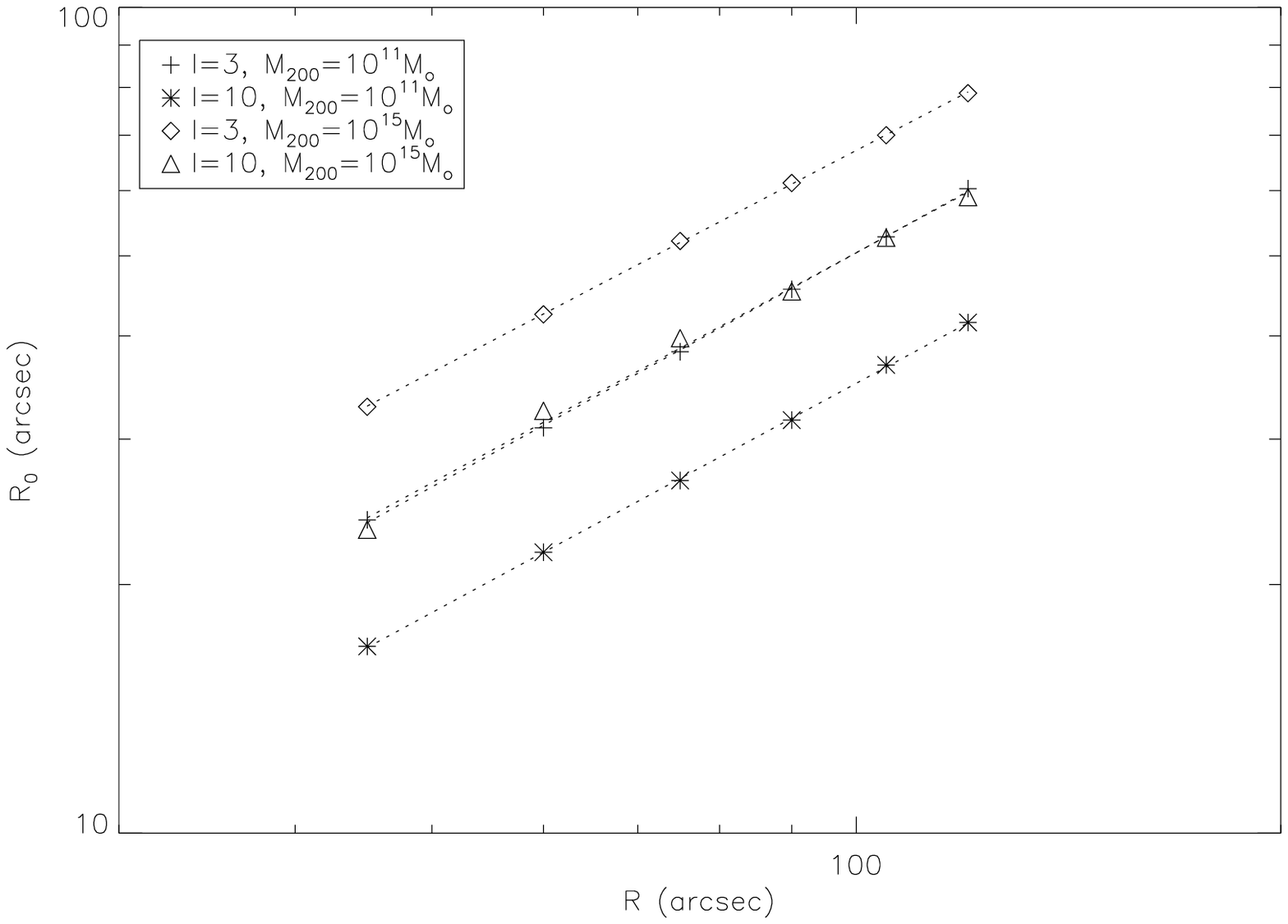}
\caption[Aperture Radius Dependence of the Peak Aperture Mass Signal and Zero-signal Radius for an NFW Lens]{Peak $\fmap$ signal (\textit{top panel}) and zero-signal radius (\textit{bottom panel}) as a function of aperture size $R$ for an NFW model with a concentration parameter of $c=3$. The dotted lines show the best fit power laws.\label{fg:nfw_r}}
\end{figure}

Figure \ref{fg:nfw_r} shows the peak signal and zero-signal radius for fixed mass, concentration parameter ($c=3$), and $l$ as a function of $R$. Shown on a logarithmic scale, the behaviour of the peak signal clearly appears to be a power law in $R$, i.e. $m_{peak}=aR^n$. From the plots shown, it is apparent that $a$ and $n$ are strongly dependent on the mass and only weakly dependent on the polynomial order, $l$. Here again, the zero-signal radius appears to follow a power law behaviour, with $R_0=bR^p$. Table \ref{tab:nfw_r} shows the fit parameters for the combinations shown in the figure. The figure shows a degeneracy in values of $R_0$ between the reconstructions with $l=10,\ M_{200}=10^{15}h^{-1}\,$M$_{\odot}$ and those with $l=3,\ M_{200}=10^{11}h^{-1}\,$M$_{\odot}$. This is to be expected: for a fixed mass, concentration parameter and aperture size, increasing $l$ will decrease $R_0$ as the filter will be narrower in width. On the other hand, for a fixed concentration parameter, increasing the mass increases the scale radius $\theta_s$, which will have the effect of increasing $R_0$ if all aperture parameters remain fixed. 

\begin{table}
\centering
\begin{tabular}{c c c c c c}
\hline
$l$ & $M_{200} (M_\odot)$ & $\log_{10}(a)$ & $n$ & $\log_{10}(b)$ & $p$\\
\hline
3 & 10$^{11}$ & 0.4311 & 0.9576 & -0.1813 & 0.9420\\
10 & 10$^{11}$ & 0.4736 & 1.1231 & -0.3007 & 0.9227 \\
3 & 10$^{15}$ & 0.8408 & 1.8893 & 0.0041 & 0.8927 \\
10 & 10$^{15}$ & 1.0908 & 1.9300 & -0.1490 & 0.9260\\
\hline
\end{tabular}
\caption[Peak Signal and Zero-signal Radius vs $R$ Power Law Fit Coefficients for an NFW Lens]{Table of coefficients for the power law fits to the data shown in Figure \ref{fg:nfw_r}: $m_{peak}=aR^n$ and $R_0=bR^p$. Here, again, the signal is due to an NFW lens with a concentration parameter $c=3$.\label{tab:nfw_r}}
\end{table}

\begin{figure}
\center
\includegraphics[width=0.45\textwidth]{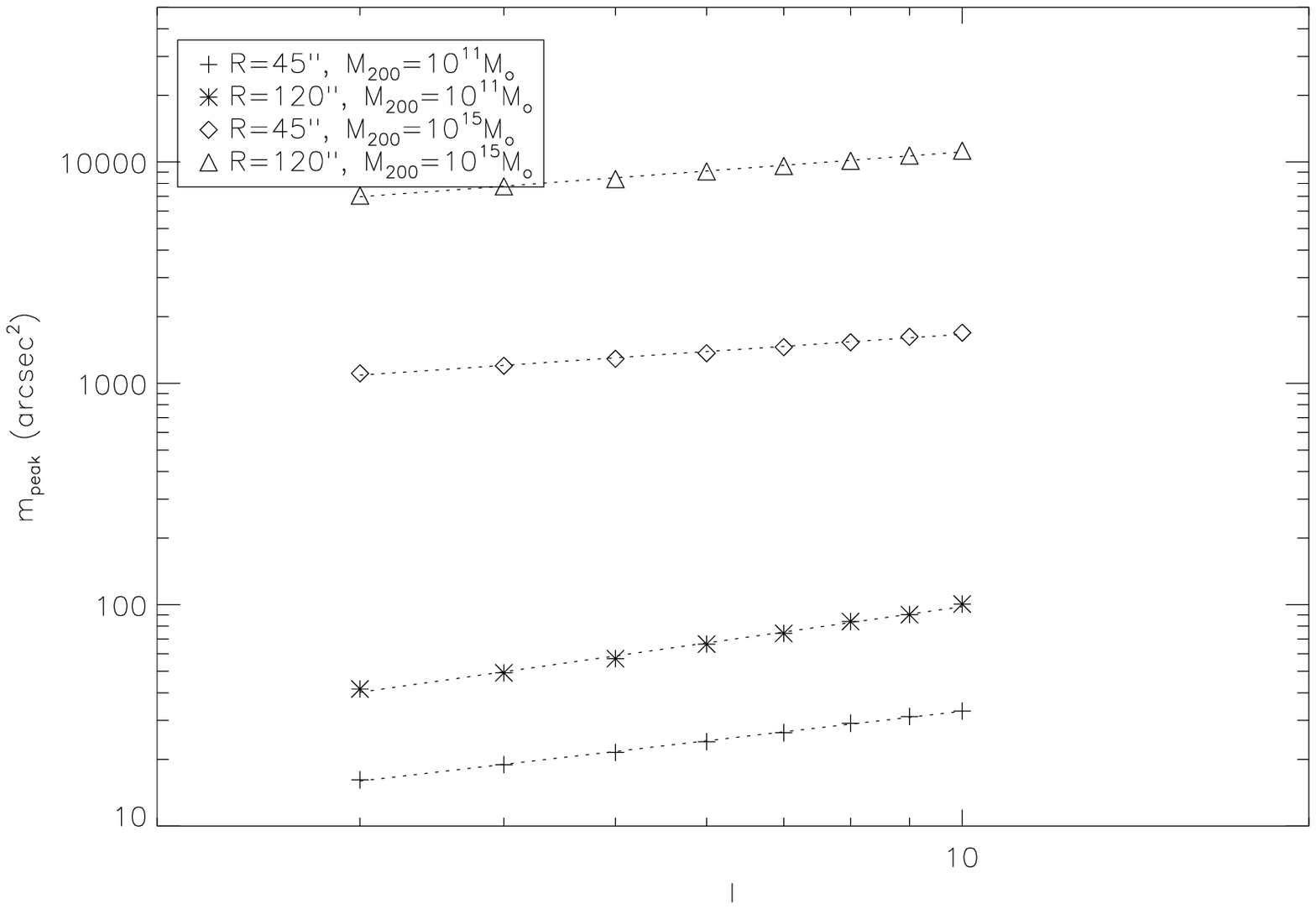}
\includegraphics[width=0.45\textwidth]{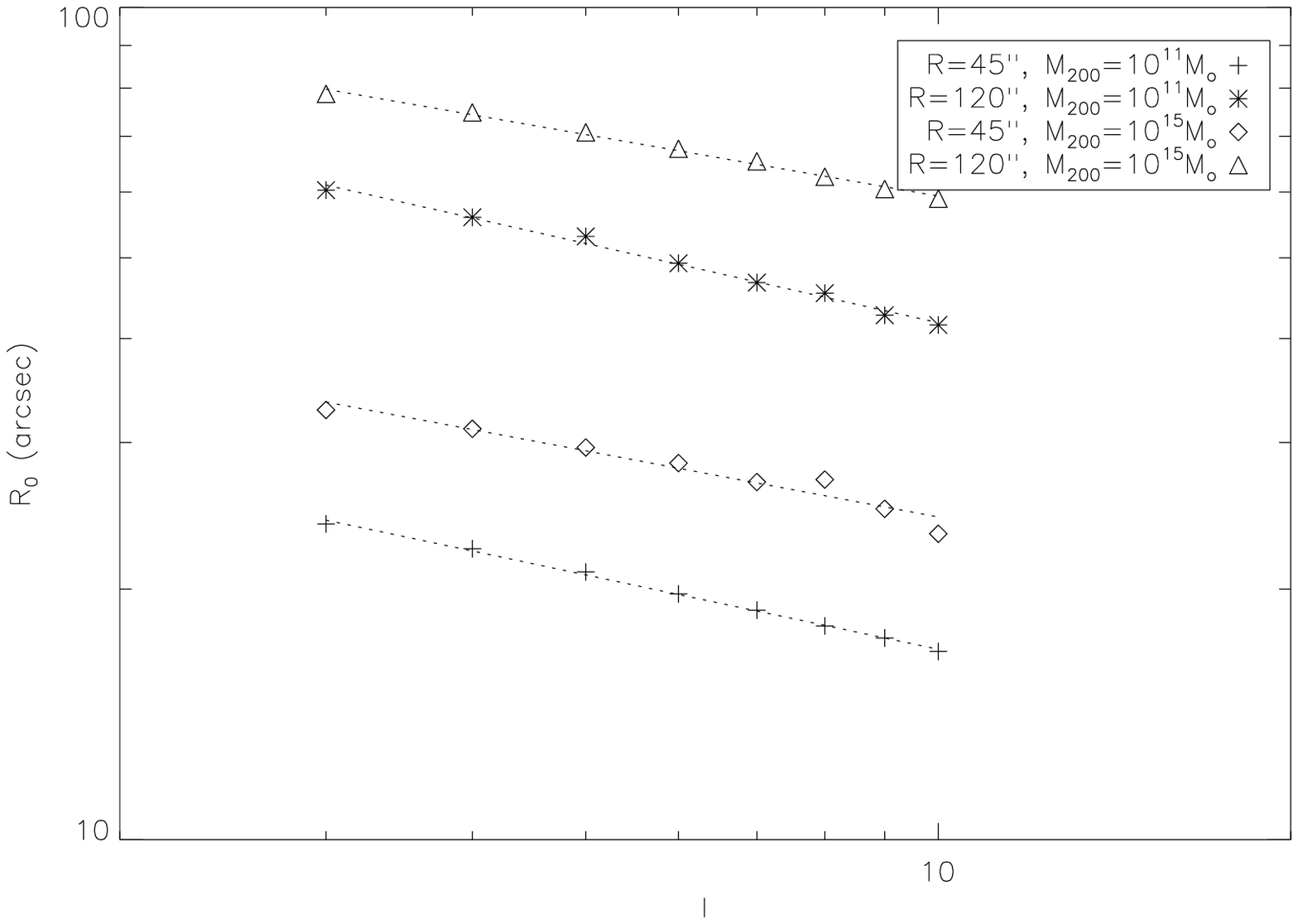}
\caption[Filter Shape Dependence of the Flexion Aperture Mass Peak Signal and Zero-signal Radius for an NFW Lens]{$m_{peak}$ (\textit{top panel}) and $R_0$ (\textit{bottom panel}) as a function of $l$ for various choices of $M_{200}$ and $R$ for an NFW lens with a concentration parameter of $c=3$. The dotted lines show the best fit power law to each set of data. The fit parameters are given in Table \ref{tab:nfw_l}.}
\label{fg:nfw_l}
\end{figure}

We now consider the behaviour of the peak signal and zero-signal contour as a function of $l$ at fixed mass and $R$. Figure \ref{fg:nfw_l} shows this behaviour for various values of $M_{200}$ and $R$, with $c$ again fixed at 3. Again, this behaviour seems well fit by a power law in $l$ with variable index; i.e. $m_{peak}=al^n$ and $R_0=bl^p$. As expected, we see that increasing $l$ while holding all other parameters fixed decreases the zero-signal radius. Table \ref{tab:nfw_l} shows the best fit parameters for this power law for the data shown in Figure \ref{fg:nfw_l}.

\begin{table}
\centering
\begin{tabular}{c c c c c c}
\hline
$R(''$) & $M_{200}(M_\odot)$ & $\log_{10}(a)$ & $n$ & $\log_{10}(b)$ & $p$\\
\hline
45 & $10^{11}$ & 0.9151 & 0.6039 & 1.5255 & -0.2968\\
120 & $10^{11}$ & 1.2521 & 0.7390 & 1.9375 & -0.3163\\
45 & $10^{15}$ & 2.8683 & 0.3537 & 1.6518 & -0.2641\\
120 & $10^{15}$ & 3.6576 & 0.3876 & 2.0187 & -0.2454\\
\hline
\end{tabular}
\caption[Peak Signal and Zero-Signal Radius vs $l$ Power Law Fit Coefficients for an NFW Lens]{Best fit parameters for the data shown in Figure \ref{fg:nfw_l} assuming the relationships to be of the form $m_{peak}=al^n$ and $R_0=bl^p$. The lens in this case is an NFW lens with a concentration parameter of $c=3$.\label{tab:nfw_l}}
\end{table}

\begin{figure*}
\center
\includegraphics[width=0.45\textwidth]{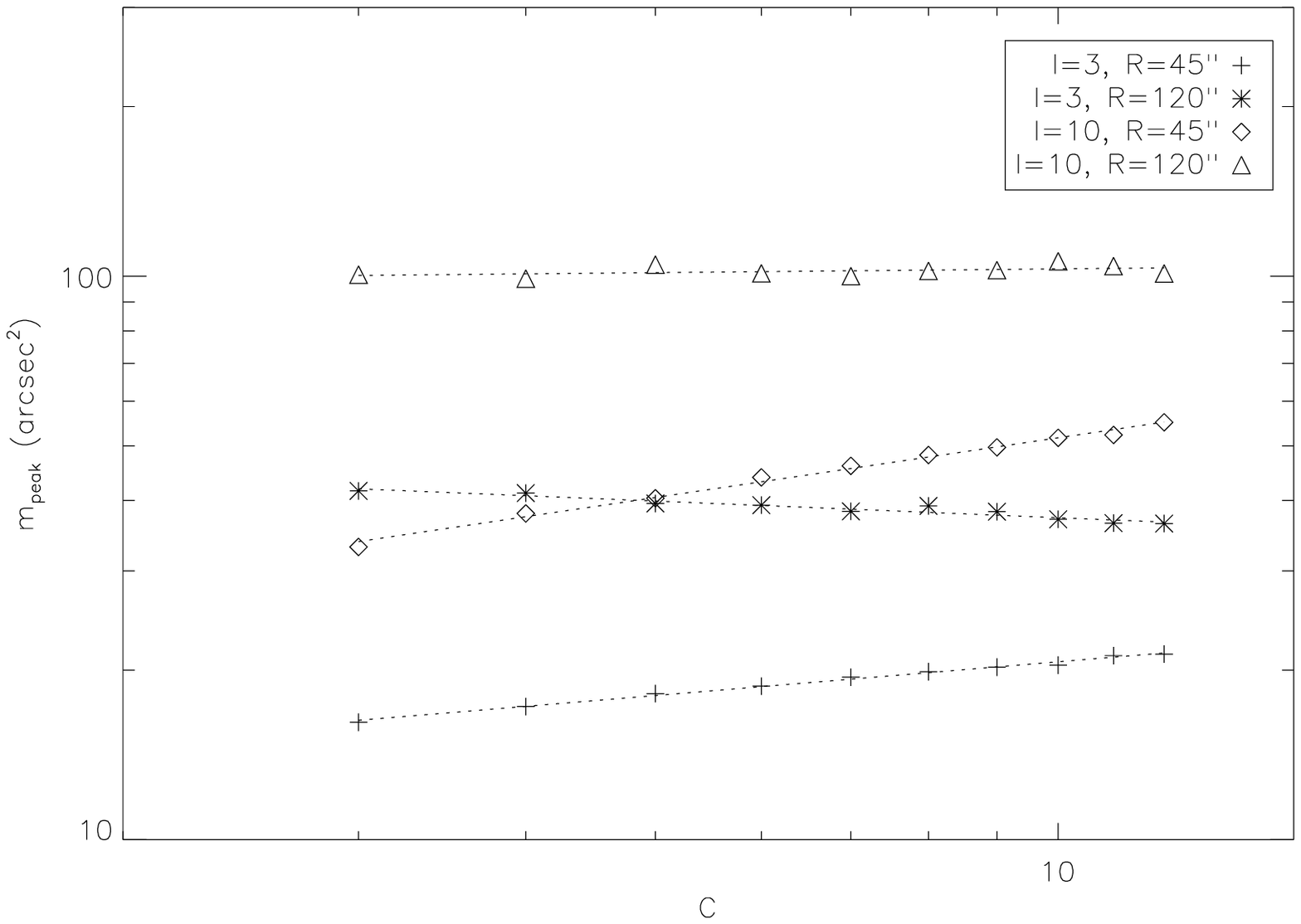}\includegraphics[width=0.45\textwidth]{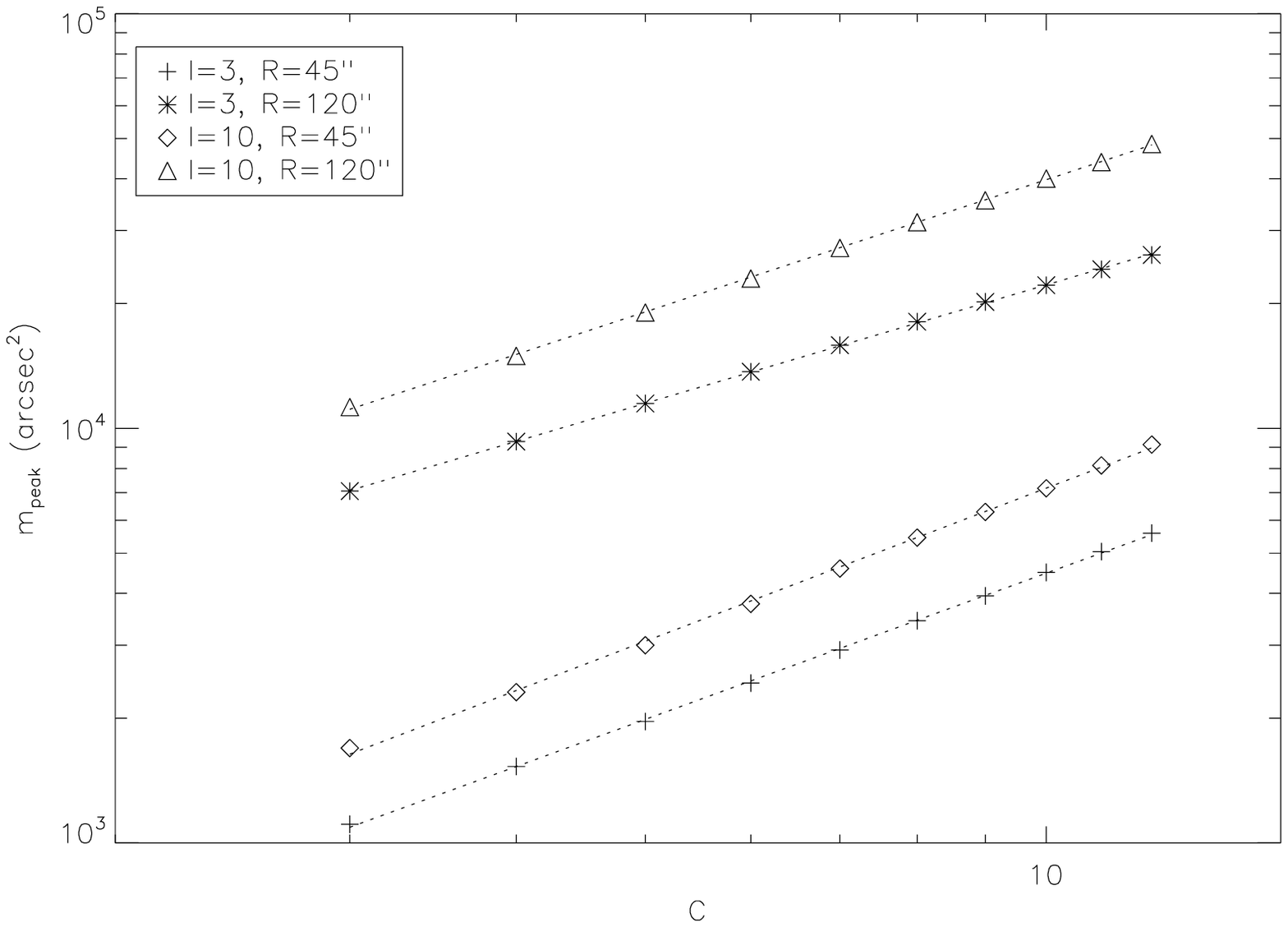}
\includegraphics[width=0.45\textwidth]{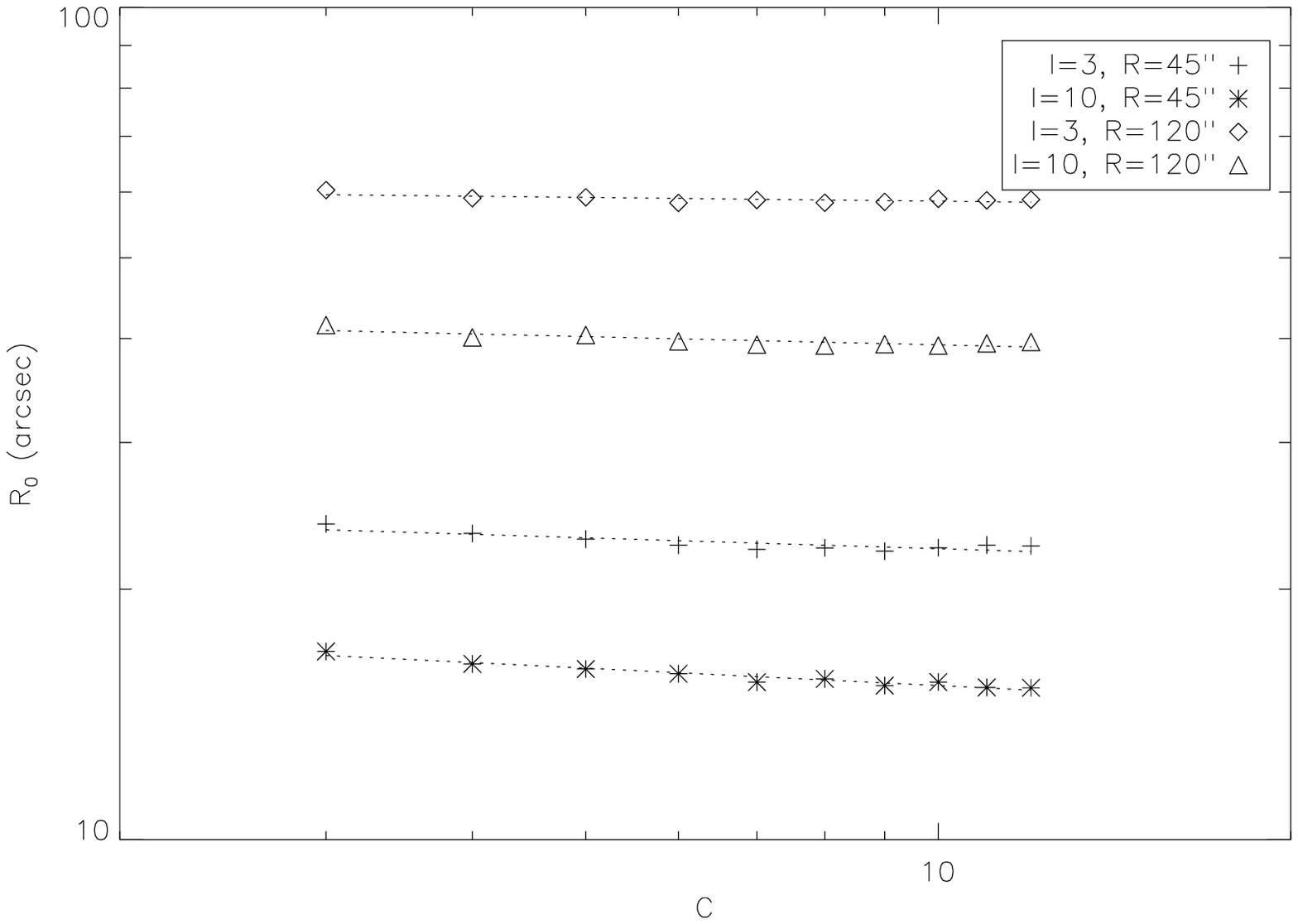}\includegraphics[width=0.45\textwidth]{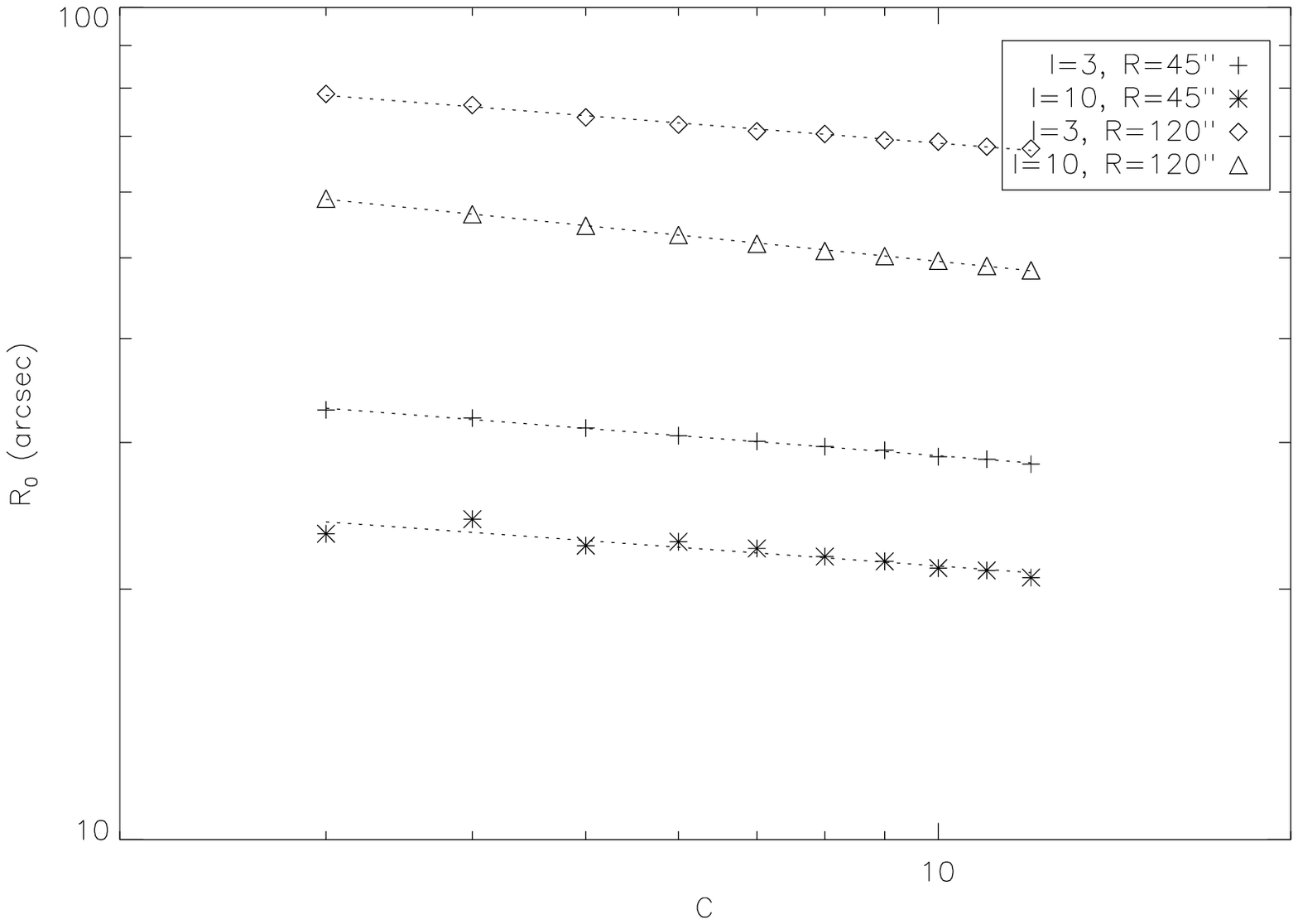}
\caption[Concentration Parameter Dependence of the Aperture Mass Peak Signal and Zero-Signal Radius for an NFW Lens]{The figure shows the behaviour of the peak $\fmap$ signal (\textit{top row}) and the zero-signal radius (\textit{bottom row}) as a function of the NFW concentration parameter for fixed values of $l$ and $R$ for a low mass halo (\textit{left panel}, $10^{11}h^{-1}\,M_\odot$), and a high mass halo (\textit{right panel}, $10^{15}h^{-1}\,M_\odot$). The dotted lines show the best fit power law to the data, the fit parameters of which can be found in Table \ref{tab:nfw_c}.\label{fg:nfw_c}}
\end{figure*}

Finally, we consider the behaviour of the peak $\fmap$ signal and the zero-signal radius for an NFW profile of fixed mass, with fixed aperture parameters $R$ and $l$, as a function of concentration parameter $c$. The data have a strong dependence on halo mass, therefore Figure \ref{fg:nfw_c} shows the behaviour of the peak signal and zero-signal contours for various combinations of $l$ and $R$ for both a halo of mass $10^{11}h^{-1}\,M_\odot$ and $10^{15}h^{-1}\,M_\odot$. Here, again, the behaviour of each appears to be a power law of the form $m_{peak}=ac^n$ and $R_0=bc^p$. Table \ref{tab:nfw_c} shows the best fit parameters for these power laws for the data shown in Figure \ref{fg:nfw_c}.

\begin{table*}
\centering
\begin{tabular}{c c c c c c c}
\hline
$M_{200}(M_\odot)$ & $l$ & $R('')$ & $\log_{10}(a)$ & $n$ & $\log_{10}(b)$ & $p$\\
\hline
$10^{11}$ & 3 & 45 &  1.1171 & 0.1986 & 1.3931 & -0.0435\\
$10^{11}$ & 3 & 120 & 1.6694 & -0.0980 & 1.7821 & -0.0149\\
$10^{11}$ & 10 & 45 &  1.3612 & 0.3521 & 1.2545 & -0.0693\\
$10^{11}$ & 10 & 120 & 1.9908 & 0.0222 & 1.6273 & -0.0328\\
$10^{15}$ & 3 & 45 &  2.4788 & 1.1720 & 1.5705 & -0.1092\\
$10^{15}$ & 3 & 120 & 3.3976 & 0.9483& 1.9467 & -0.1098\\
$10^{15}$ & 10 & 45 &  2.6293 & 1.2265 & 1.4302 & -0.1013\\
$10^{15}$ & 10 & 120 & 3.5409 & 1.0588 & 1.8376 & -0.1427\\
\hline
\end{tabular}
\caption[Peak Signal and Zero-signal Radius vs $c$ Power Law Fit Coefficients for an NFW Lens]{Best fit parameters for the data shown in Figure \ref{fg:nfw_c} assuming relationships of the form $m_{peak}=ac^n$ and $R_0=bc^p$.\label{tab:nfw_c}}
\end{table*}

The behaviour of the peak $\fmap$ signal from the NFW profile differs greatly from that of the SIS model as the aperture radius and filter polynomial order are changed. Moreover, the behaviour changes with halo mass and concentration parameter in a non-trivial way. A similar trend is seen when considering the behaviour of the zero-signal radius, where the slope of the power law behaviour changes based on aperture size, filter shape, virial mass and concentration parameter. Again, it is not possible to model this behaviour simply and arrive at a general analytic expression for the zero-signal radius that simultaneously describes its behaviour as a function of all the parameters one might vary. 

This vastly differing behaviour 
implies that if one were to consider several flexion aperture mass reconstructions of a lens field using different aperture radii and filter polynomial order, one might be able to distinguish not only between an SIS model and an NFW model, but between NFW models with different masses and concentration parameters. Moreover, considering the behaviour exhibited in Figures \ref{fg:nfw_r} and \ref{fg:nfw_l}, it appears that a change in aperture radius shows a greater change in the overall peak signal than a change in polynomial order, and thus might provide a better discriminant between profiles than the filter polynomial order.

\section{Model-selection with $\fmap$}
\label{sec:discrim}

Having characterised the expected behaviour of the peak signal and zero-signal radius of the $\fmap$-statistic for the SIS model and a family of NFW models of varying concentration parameters, one might well ask how the differences seen between the various models might be used to constrain lens parameters in real observations. The first key feature seen in \S~\ref{sec:radprof} is that the value of the peak signal is a strong function of the virial mass of the lens, regardless of which mass model one is considering. This implies that a measurement of the peak $\fmap$ signal will automatically, at very least to within an order of magnitude, give an indication of the mass of the structure responsible for the lensing signal. 

Further, the data above demonstrate that, under a change in aperture radius, the peak signal and zero-signal radius change in different ways for each of the two models, and a strong variation is also seen between NFW models with different concentration parameters. Such divergent behaviour is also seen when varying the filter polynomial order, $l$, though to a lesser extent. The simulated data therefore suggest that given a set of $\fmap$ data for various combinations of $l$ and $R$, the mass and mass profile of a halo might be determined uniquely by considering the behaviour of the peak signal and zero-signal radius as $l$ and $R$ are varied. In other words, $\fmap$ filtering of the flexion signal for a range of aperture parameters allows one to delineate between various mass models, offering a method by which degeneracies between parametric models might be broken.

In order to assess this further, we utilise the simulated radial profiles described in \S~\ref{sec:radprof} above. These simulations were evaluated for $M_{200}/(10^{11}h^{-1}\,M_\odot) = $ 1, 2, and 5 $\times 10^{0}$, $10^{1}$, $10^2$ and $10^3$. For each mass, an SIS profile was evaluated, as well as NFW profiles with $c$ sampled in integer values in the range $3\le c\le12$. Further, the $\fmap$ signal for each model was evaluated using all possible combinations of $l=[3,4,5,6,7,8,9,10]$ and $R=[45'',60'',75'',90'',105'',120'']$. In order to improve the resolution attainable at the low-mass end, additional simulations were carried out for the SIS and all the above NFW profiles with $M_{200}=[3,4,6,7,8,9]\times10^{11}h^{-1}\,M_\odot$ and all combinations of $l=[3,5,7,10]$ and $R=[60'',90'',120'']$. Similarly, at the high-mass end, we added simulations involving $M_{200}=[3,4,6,7,8,9,20]\times10^{14}h^{-1}\,M_\odot$ and all combinations of $l=[3,5,7,10]$ and $R=[45'',60'',75'',90'',120'']$. 

The discriminating power of the $\fmap$-statistic is tested as follows. We consider a lens of a given mass $M_{200}^\ast$ and concentration parameter $c^\ast$. We assume that we are able to measure the $\fmap$-statistic for a number of combinations $[l,R]$ with a signal to noise of ${\cal S}$ and an error on the measurement of $R_0$ given by $\sigma_{Ro}$. For each combination of $l$ and $R$, we find the expected value of $m_{peak}^\ast$ and $R_0^\ast$ for our test model from the simulations already carried out. We then scan the parameter space to find any other models with values of $m_{peak}$ and $R_0$ that fall within our error bars for the combination of $l$ and $R$ being used; i.e. $m_{peak}^\ast-\sigma_{peak}\le m_{peak} \le m_{peak}^\ast+\sigma_{peak}$ and $R_0^\ast-\sigma_{Ro}\le R_0\le R_0^\ast+\sigma_{Ro}$ (note that $\sigma_{peak}=m_{peak}^\ast/{\cal S}$). As $l$ and $R$ are changed, the lists of models that fit the test model's ``data'' within the error bars are compared, and only those models found to be compatible with \textit{all} the ``measurements'' carried out using our model lens are retained.

\subsection{Galaxy-Mass Halo}

As a first test of this method, we consider a theoretical lens with a mass of $M_{200}^\ast=5\times10^{11}h^{-1}\,M_\odot$ with an NFW profile with concentration parameter $c^\ast=12$. We assume that we are able to measure the $\fmap$-statistic for $l=[3,5,7,10]$ and $R=[60'',90'',120'']$ with a modest signal to noise of 2.0 and an error on the measurement of $R_0$ of $\pm1''$. After iterating over all the available combinations of $l$ and $R$, the only models in the simulated data set that are found to have consistently fallen within the error bars on the theoretical measurements are: $M_{200}=4\times10^{11}h^{-1}\,M_\odot$ with $c=7,\ 8,\ {\rm or}\ 12$,  $M_{200}=5\times10^{11}h^{-1}\,M_\odot$ with $c=10,\ 11,\ {\rm or}\ 12$, and  $M_{200}=7\times10^{11}h^{-1}\,M_\odot$ with $c=12$. If the error on the measurement of $R_0$ is reduced to $\pm0.5''$, the degeneracy is completely broken and we recover our input model, even for a signal to noise of 1. Note that these results are typical for this mass range. Changing the input value of $c^\ast$, or using an SIS profile, showed a similar success rate. 

We can conclude therefore that this method is successful at recovering both the total mass and input concentration parameter to within a factor of $<1.5$, and effective at eliminating alternative parametrisations for the mass profile such as the SIS model or NFW models with significantly higher or lower concentration parameters, given a sufficient number of $\fmap$ realisations. Moreover, in the low-mass case, reduction in the errors on measurements of $R_0$ are most effective at reducing the number of models compatible with the data, and thus the error on the estimates of the mass and concentration parameter.

\subsection{Cluster-Mass Halo}

We now consider a lens with a mass $M_{200}^\ast=5\times10^{14}h^{-1}\,M_\odot$ and concentration parameter $c^\ast=4$, and consider measurements at $l=3,\ R=[60'', 90'', 120'']$ with ${\cal S}=2$ and $\sigma_{Ro}=1''$. After only three iterations, the field of possible models has been narrowed down to $M_{200}=2\times10^{14}h^{-1}\,M_\odot$ with $c=3$, $M_{200}=[4,5,6]\times10^{14}h^{-1}\,M_\odot$ with $c=4$ and $M_{200}=7\times10^{14}h^{-1}\,M_\odot$ with $c=5$. Increasing the signal to noise to $2.5$ isolates models with $c=4$ and $M_{200}$ in the range $4\times10^{14}h^{-1}\,M_\odot\le M_{200}\le 6\times10^{14}h^{-1}\,M_\odot$. In this case, reducing the error on $R_0$ to $0.5''$ does not completely break the degeneracy, but does reduce the number of available models to 2, finding $M_{200}=4-5\times10^{14}h^{-1}\,M_\odot$ with $C=4$. However, increasing the number of models iterated over does break the degeneracy for modest error estimates. Using $l=3$ and $R=[45'',60'',90'',120'']$ with ${\cal S}=2.5$ and $\sigma_{Ro}=1.0$ returns our original model. 

In the case of high mass haloes, far fewer $\fmap$ realisations are required for convergence than in the low mass case. The data here suggest that with only four $\fmap$ reconstructions of a cluster-mass halo and reasonable errors on the zero-signal radius and the peak signal to noise allow us to recover our input model uniquely up to the resolution of our sample data set. In this case, however, greater benefit is obtained by increasing the peak signal to noise in the $\fmap$ reconstruction or increasing the number of reconstructions obtained, rather than decreasing the errors on the measurement of $R_0$.

\section{Test on a Simulated Cluster}
\label{sec:nfw}

In order to more realistically evaluate the performance of this method, we employ the $\fmap$ data from LKW09 for a simulated cluster taken from the Millennium simulation (Springel et al., 2004). This cluster, on extraction from the simulation, was found to have a virial mass of $M_{200}^\ast=1.23\times10^{15} h^{-1}\,M_\odot$, and is located at a redshift of $z_{\rm d}=0.21$. Background sources were generated at random positions with a density of $35\,$arcmin$^{-2}$ at a redshift of $z_{\rm s}=1.0$. The background sources were assumed to have zero intrinsic flexion, and an ellipticity drawn randomly from a Gaussian distribution with a standard deviation of $\sigma_\epsilon=0.2$. 

These galaxies were lensed through the input cluster potential using a ray-tracing procedure (see LKW09 for details of this calculation). As the galaxies had zero intrinsic flexion, the dominant sources of noise in the reconstructions arise from measurement errors and errors in the ray-tracing code resulting from the high degree of pixellation of the input convergence, which was not smoothed. As a result, one would expect a somewhat lower signal to noise to be seen in real data assuming the same density of background sources.

\subsection{Direct Nonparametric Mass Reconstruction}

\begin{figure}
\center
\includegraphics[width=0.45\textwidth]{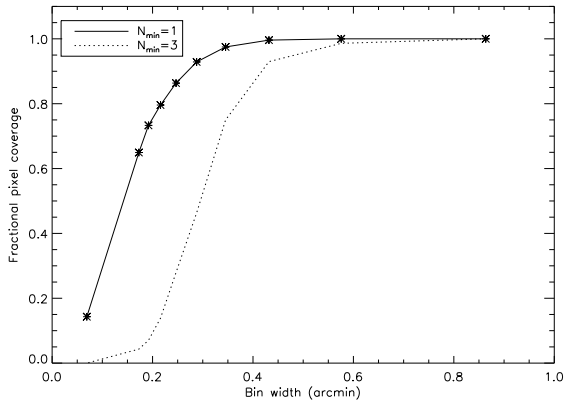}
\caption[Fractional pixel coverage]{The fraction of pixels containing at least $N_{\min}$ sources as a function of pixel size.}
\label{fg:frac_covered}
\end{figure}

When carrying out a direct reconstruction, the first issue to be addressed is that of the appropriate choice of bin size over which to average the flexion signal. A balance must be struck between using a fine grid, in order to resolve structures on both small and large scales, and having a statistically significant number of background sources within each bin. In Figure \ref{fg:frac_covered}, we show the fraction of pixels containing at least $N_{\rm min}$ sources as a function of bin width for $N_{\rm min}=1$ and $3$.

To alleviate additional pixel noise in our reconstruction resulting from incomplete coverage, we choose a bin size of $0.43\,$arcmin per pixel, resulting in a reconstructed convergence map of 16 x 16 pixels. In order to reduce edge effects when carrying out the Fourier transforms, we add zero-padding on all four sides of the image such that the padded image consisted of 128 x 128 pixels. To estimate the noise in our reconstruction, the flexion vectors were rotated by a random angle, and the reconstruction repeated. This procedure was carried out 1000 times, and the error at each pixel location estimated by the standard deviation $\sigma_i$ of the randomised reconstructions.

Finally, the reconstructed convergence must be appropriately rescaled. As our data covers only the central 1$h^{-1}\,$Mpc of the cluster, we do not expect the convergence to go to zero at the edge of the cluster. We therefore set the integration constant $\kappa_0$ such that the weighted mean of the reconstructed convergence, defined as:
\begin{equation}
\overline{\kappa}=\left[\sum_{\rm pixels} \frac{\kappa_i}{\sigma_i^2}\right] \left[\sum_{\rm pixels} \frac{1}{\sigma_i^2}\right]^{-1}\ ,
\end{equation}
is equal to the mean convergence in the input map, found to be $\overline{\kappa}=0.13$.

\begin{figure}
\center
\includegraphics[width=0.45\textwidth]{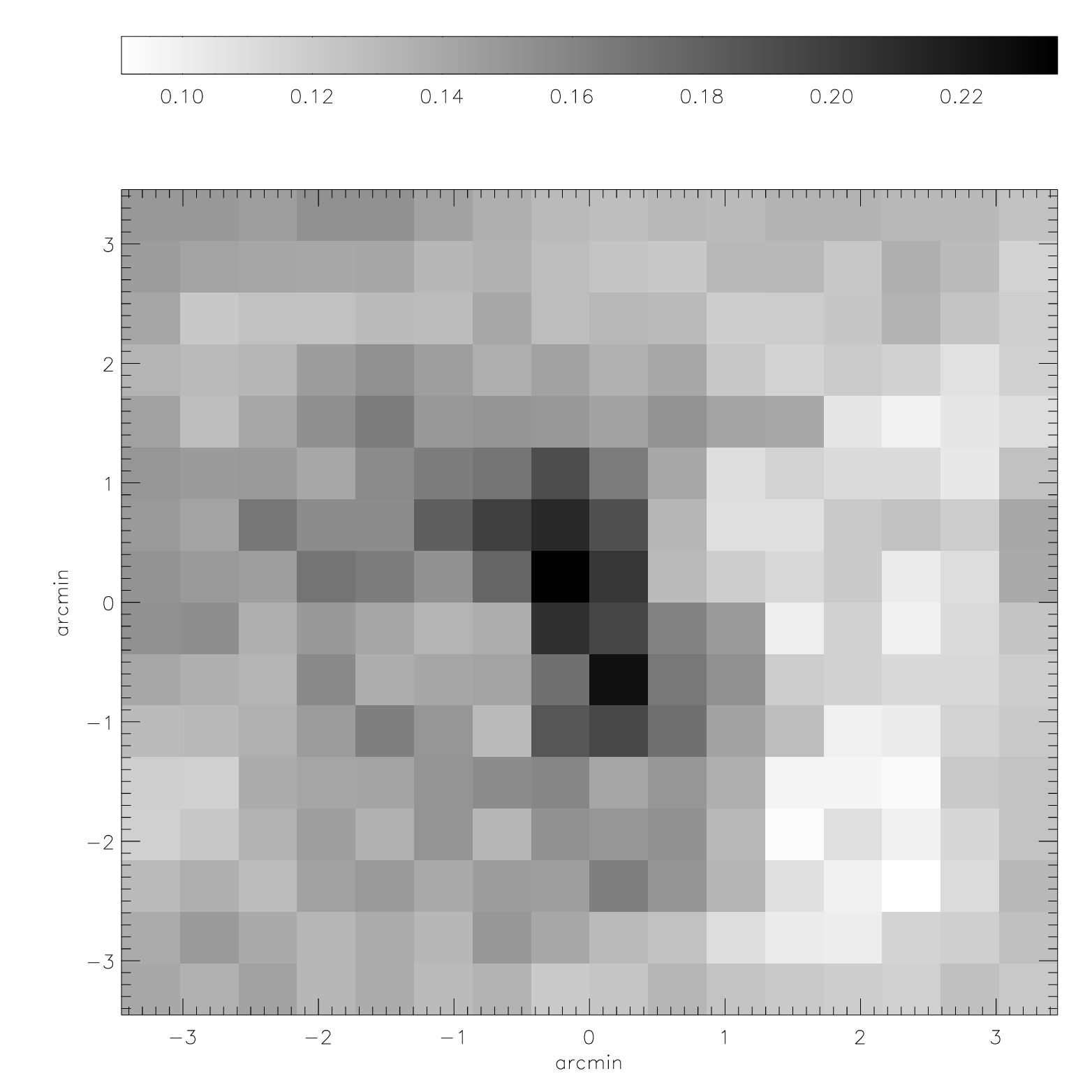}
\includegraphics[width=0.45\textwidth]{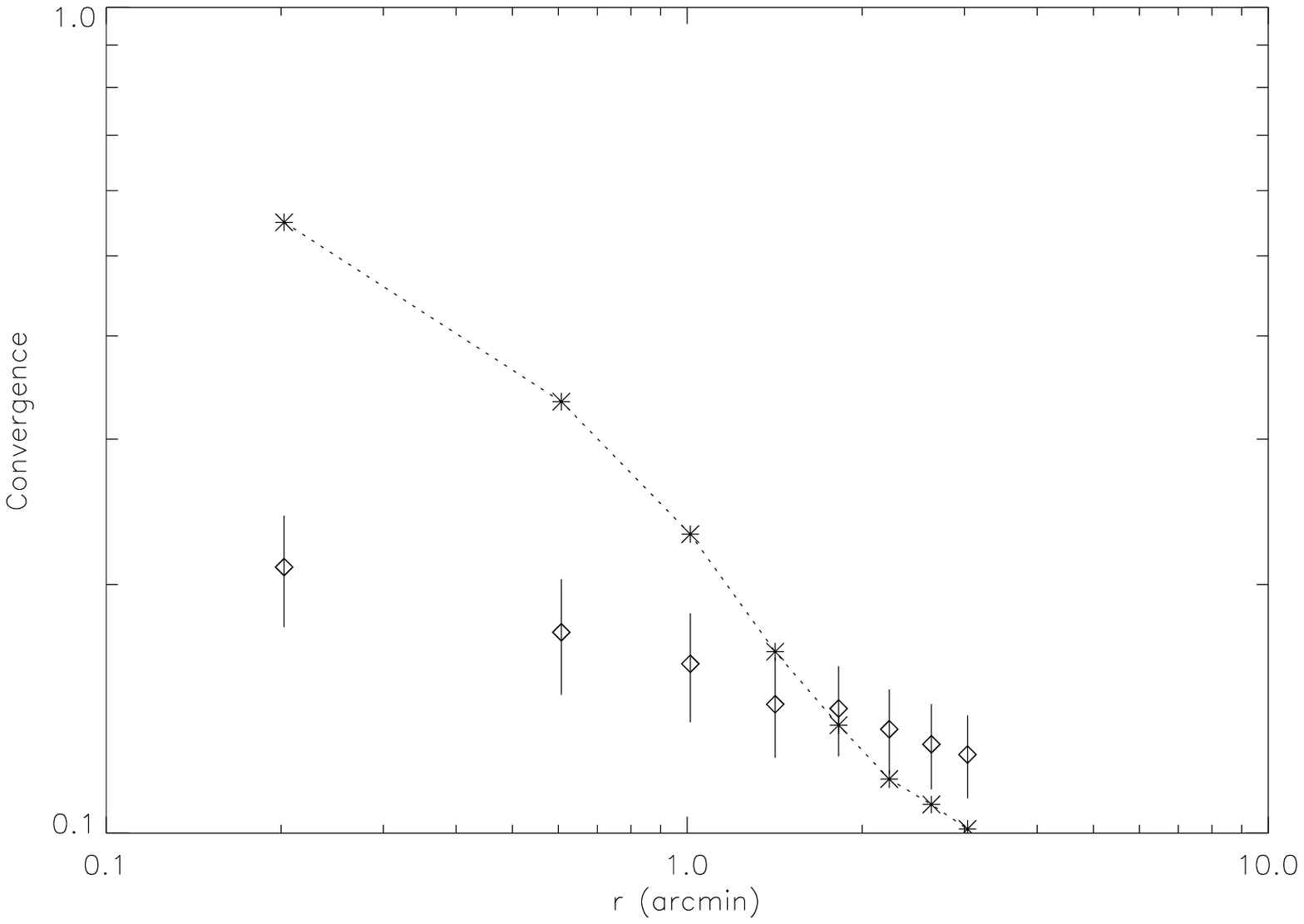}
\caption[FFT Reconstruction from Flexion Measurements]{\textit{Top Panel}: A reconstruction of the convergence map of the cluster from flexion measurements. \textit{Bottom Panel}: A comparison of the radially-averaged convergence profile of the input convergence (dotted line with points) and the reconstructed convergence (open diamonds with error bars). }
\label{fg:directrec}
\end{figure}

Figure \ref{fg:directrec} shows the convergence reconstructed in this manner, as well as a comparison of the radial profile of the reconstruction with that of the input convergence map. Note that the error bars represent the $1$-$\sigma$ errors on $\kappa-\kappa_0$. Modelling the convergence as a power law of the form $\kappa \propto r^{-\alpha}$, we find that the convergence reconstruction shows a shallow slope ($\alpha = 0.19$), while the input map shows a much steeper slope ($\alpha \sim 0.66$). We note also that the input convergence appears to follow a broken power law, rather than having a constant power law index, whilst the reconstruction does not appear to follow this behaviour.

\begin{figure}
\center
\includegraphics[width=0.45\textwidth]{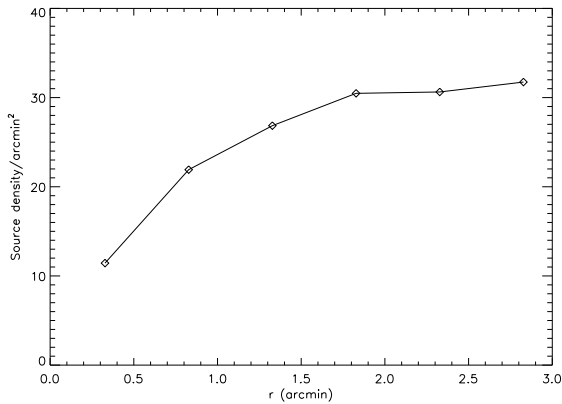}
\caption{The background sources averaged in annuli of width 0.5 arcminutes as a function of distance from the centre of the cluster.}
\label{fg:source_count}
\end{figure}

The underestimation of the central slope is not entirely unexpected. The reason is that one expects a decrease in the number of background sources with flexion measurements in the central regions of the cluster, resulting from sources being lensed away from the centre, and from blending of background images with those of cluster members.  Figure \ref{fg:source_count} shows the mean background source count evaluated in annuli as a function of distance from the cluster centre, demonstrating a significant underdensity in the central 1.5 arcminutes as compared with the periphery. The lack of sources in the central region limits the ability of this method to constrain the potential in that regime.

Note that due to our choice of normalisation constant, the integrated mass found through the direct reconstruction method will be consistent with that of the input map; however, it is clear from Figure \ref{fg:directrec} that this method will not allow one to determine with any confidence the true shape of the underlying mass profile. Moreover, the choice of normalisation constant relies either on the presence of data at large radii from the centre of the cluster, or on other external information. In the absence of either of these, the choice of normalisation, and hence the cluster mass thus derived, is somewhat arbitrary.

\subsection{$\fmap$ Analysis and Results}

As in LKW09, the $\fmap$-statistic was measured for $l=5$ and $R=[60'',90'',120'']$. Hence, to test how well this profile might be identified, $\fmap$ data for an SIS model and the full range of NFW models were obtained for $M_{200}^\ast$, and added to our sample of test profiles.

The peak signal to noise in the reconstructions was found to be ${\cal S}=2.5, 3.0,$ and $3.3$, in order of increasing aperture radius, and we assume an error on the zero-signal radius measurements of $\sigma_{Ro}=10''$, which corresponds to $\sim 5$ pixels on the input convergence map. 

We note that in each reconstruction shown in LKW09, there is a significant peak seen at the top edge of the field of view. In the $R=60''$ $\fmap$ reconstruction, this feature actually obtains higher signal to noise than the central peak. This feature is believed to arise due to edge effects, as the significance of this peak is reduced in subsequent reconstructions, and its position changes somewhat as the filter is changed. This indicates a noise feature, rather than a true signal. As we are considering a radially-averaged signal here centred on the centre of the $\fmap$ field, and this particular feature is located far outside the zero-signal radius of the central mass peak, this feature will not contaminate our measurements substantially.

\begin{figure}
\center
\includegraphics[width=0.45\textwidth]{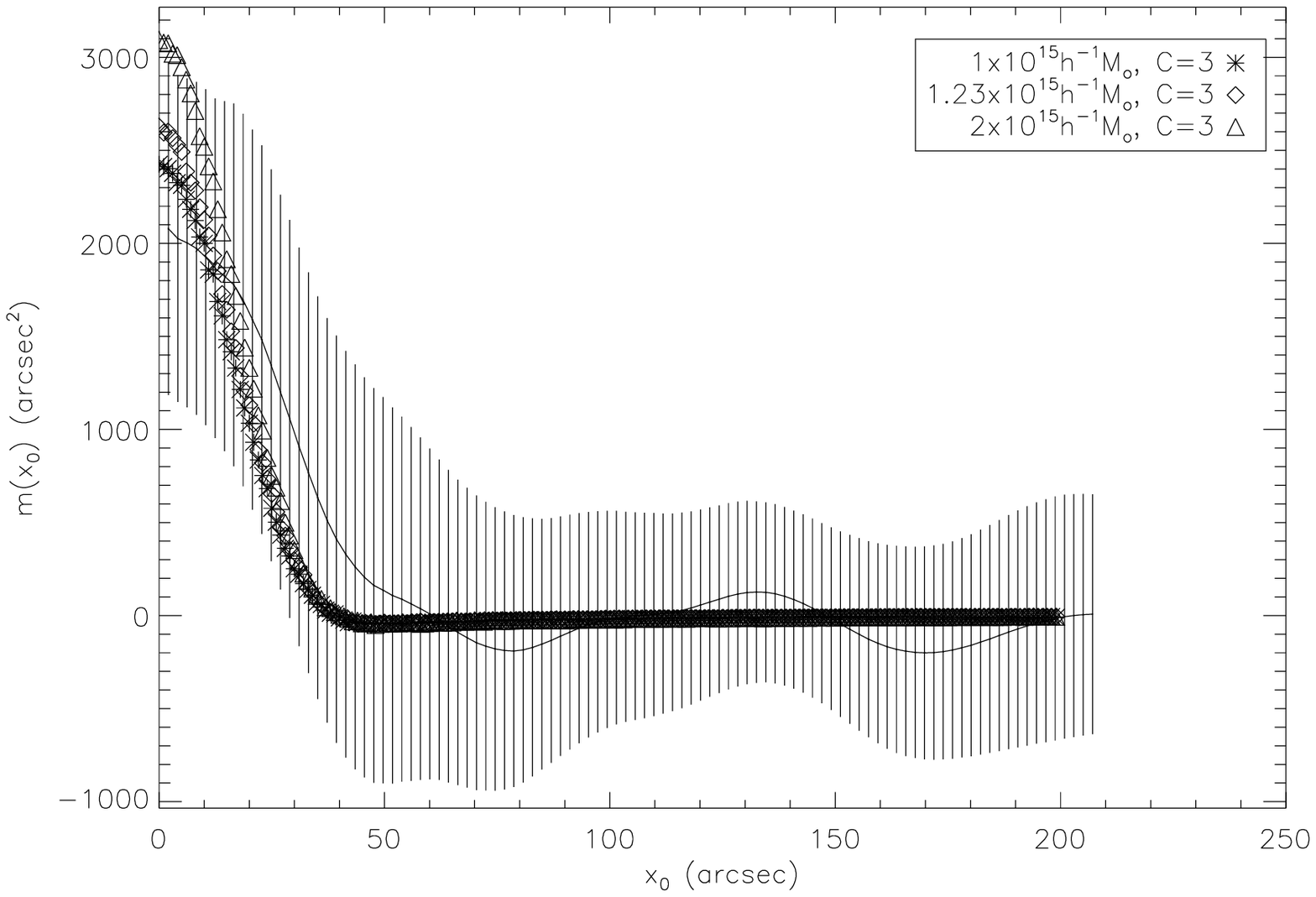}
\includegraphics[width=0.45\textwidth]{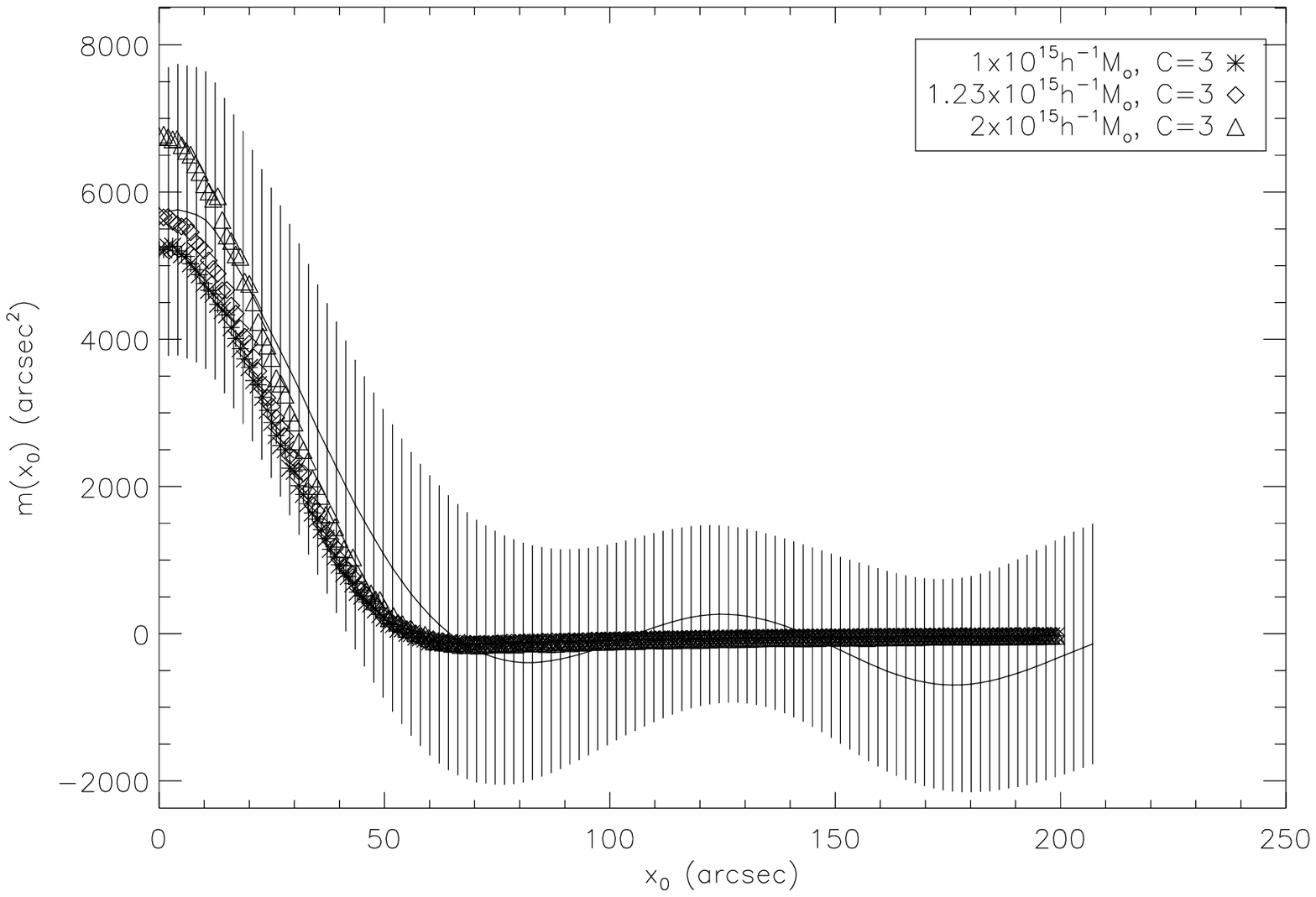}
\includegraphics[width=0.45\textwidth]{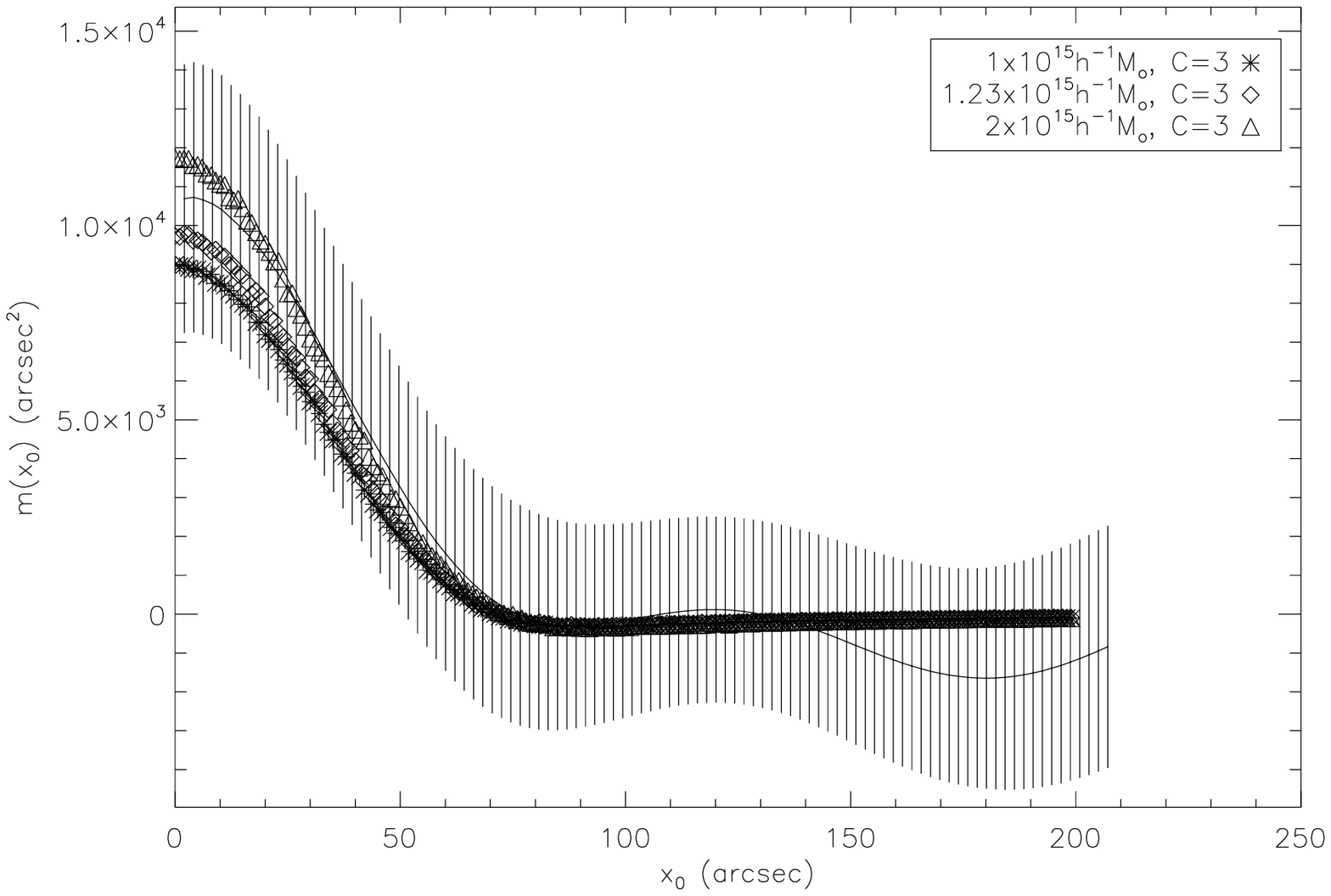}
\caption[$\fmap$ Signal from Simulated Cluster Lens]{The figure shows the radially-averaged $\fmap$ signal for the N-body cluster described in LKW09 (solid curves), and the simulated data corresponding to the two best fit models from the simulations carried out in this paper (discrete points) for $l=5$ and $R=60''$ (\textit{top panel}), $90''$ (\textit{middle panel}) and $120''$ (\textit{bottom panel}).\label{fg:radnbody}}
\end{figure}

Using the three $\fmap$ reconstructions of the N-body cluster, the data indicate that the radial profile of the cluster is well-fit by an NFW mass profile with $c=3$ and $M_{200}=1,\ 1.23\ {\rm or}\ 2\times 10^{15}h^{-1}\,M_\odot$. The radial profiles for the $\fmap$ signal from the cluster for each aperture radius are plotted in Figure \ref{fg:radnbody}, along with the radial profiles of the three best-fitting models. The effect of the cluster's ellipticity and associated substructure can be clearly seen in its radial profile, with a secondary peak being seen around $x_0=130''$ in each reconstruction.

All three models show values of $R_0$ that are smaller than that seen in the cluster data, implying that the best fit model might be one with an even smaller concentration parameter (see Figure \ref{fg:nfw_c}). However, the peak signal does appear to be best fit across the three reconstructions by the profile with $M_{200}=1.23\times10^{15}h^{-1}\,M_\odot$, as expected. 

\begin{figure}
\center
\includegraphics[width=0.45\textwidth]{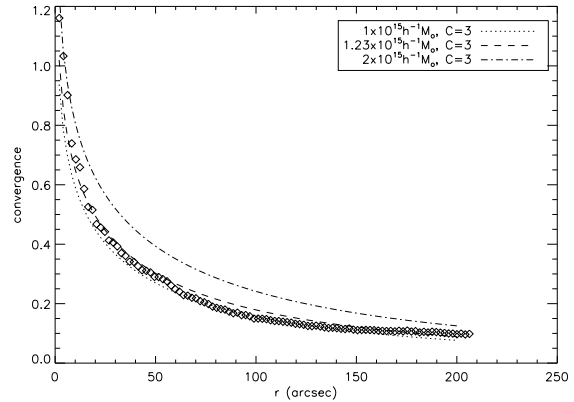}
\caption[Radial Convergence Profile for N-body Cluster]{The circularly averaged profile of the convergence of the N-body lens described in the text. The dotted curves shows the expected profile for an NFW lens with $c=3$ for various lens masses as illustrated in the legend. \label{fg:kapparad}}
\end{figure}

Furthermore, we can compare this best-fit models to the radial profile of the convergence map. Figure \ref{fg:kapparad} shows the measured radial profile of the convergence of the lens. The dotted curves show the convergence profile for the three best-fitting mass models described above. Clearly the two lower-mass models offer excellent fits to the cluster data, thus demonstrating the power of this method to accurately characterise the mass profile of a lens using a relatively small number of $\fmap$ reconstructions.  

\section{Discussion}
\label{sec:summary}

In this paper, we have presented the expected $\fmap$ signal from two common mass profiles: the Singular Isothermal Sphere model and the Navarro-Frenk-White model. This signal is characterised by a peak value, found when the aperture location coincides with a mass concentration, and a zero-signal contour. By simulating radial $\fmap$ profiles for a large number of combinations of halo mass, NFW concentration parameter, aperture radius and filter polynomial order, the behaviour of the peak signal and zero-signal contour has been characterised for both the SIS and NFW density profiles under the assumption of circular symmetry.

The behaviour of both these measures is seen to diverge rather sharply between different models with the same mass as the aperture properties are varied. This implies that one might be able to use the $\fmap$-statistic to constrain the masses and mass profiles of structures detected simply by varying the aperture parameters and noting the change in the peak signal and the location of the zero-signal contour.

Indeed, it has been shown using simulated data from analytic models that with modest signal to noise and relatively large errors on the measurement of the zero-signal radius, it is possible to isolate an input model using a reasonable number of combinations of aperture size and filter polynomial order. It also becomes easier to discriminate between masses at the high-mass end of the spectrum considered here, with higher mass haloes requiring fewer combinations of $l$ and $R$ to reach convergence. 

This was further reinforced by testing the method on $\fmap$ data using the simulated cluster of LKW09 as the input lens. This lens has a virial mass of $1.23\times10^{15}h^{-1}\,M_\odot$, and the combination of three $\fmap$ measurements of the lens was able to recover this mass to within a factor of $<1.5$, and to correctly identify the shape of the profile as an NFW with low concentration parameter. This success, despite the circularisation of a clearly elliptical lens profile, demonstrates solidly the power of the $\fmap$-statistic to discriminate between haloes of different masses. 

Moreover, this method shows a dramatic improvement over direct nonparametric reconstruction techniques. Firstly, the resolution that one can obtain using a direct method is limited to a far greater extent than aperture methods. This is because the number of bins one can use is limited by the density and distribution of background sources. We find, for the simulated cluster, a best resolution of just under half an arcminute. This coarse resolution is not problematic with the simulated cluster presented here, as it has a fairly smooth large-scale distribution. However, where there are significant substructures (such as in Abell 1689), averaging a signal over large bins has the effect of washing out the signal from smaller structures. 

Aperture measures, on the other hand, filter the signal through a large aperture ($R>0.5$ arcminutes, in general), but are able to obtain fine resolution in the output aperture mass by spacing the apertures close together. Thus, finer structure can be resolved using these methods.

In addition, we have shown that the ability of direct reconstruction techniques to accurately characterise the shape of the central density profile in clusters is very limited, resulting from a decrease in the number of background sources with flexion measurements in the central regions of the cluster. Since direct methods rely on the presence of reliable flexion measurements within each pixel, the ability to measure the central density profile is generally rather limited. Again, as aperture measures involve filtering the signal from a large area, they are less limited by a dearth of sources in the central regions.

Lastly, whilst the direct method needs to be adjusted by a somewhat arbitrary constant shift, the aperture mass statistic is, by construction, independent of this normalisation.

There are often found to be somewhat large differences seen in mass estimates of clusters of galaxies found using strong lensing, weak lensing, or a combination of the two. The flexion aperture mass statistic offers a method to break the degeneracy between these models, as it clearly discriminates between different masses and mass models. As a result of filtering the data, different mass models exhibit significantly different behaviours. Thus, this method offers a unique and novel model-selection method that is complementary to other techniques currently used, and requires no prior assumptions about the underlying mass distribution.

The issue of finding the best-fit mass profile is important to cosmology. Cosmological simulations suggest dark matter haloes should follow a universal (NFW) density profile with concentration parameter anti-correlated with mass. There is much discussion regarding whether observations support these predictions. Thus, the ability to break degeneracies between models is important to cosmology; in this paper, we offer a clear prescription for a method by which this can be achieved.

There is much work remaining to be done, of course. A more finely-sampled parameter space would allow for increased accuracy in model selection, improving the signal to noise in the measurements (by improving the filter functions used, for example) would increase the discriminating power of the $\fmap$-statistic, and considering the expected radial signal from elliptical lenses or models involving substantial substructure would provide a more realistic sample of model templates from which to select. However, the work presented in this paper represents an excellent first step, and highlights the promise that the aperture mass statistic from flexion shows to be able to discriminate between masses, mass profiles and NFW concentration parameters using only a few combinations of aperture parameters. 

\section{acknowledgments}
We thank Thomson Nguyen, Neal Jackson and Antony Lewis for useful discussions, and John Helly and Ian McCarthy for providing the Millennium Simulation cluster particle data. We also thank the anonymous referee for helpful suggestions. AL is supported by a BP/STFC Dorothy Hodgkin Postgraduate Award and LJK is supported by a Royal Society University Research Fellowship.

\appendix
\label{lastpage}


\begin{thebibliography}{99}

\bibitem[\protect\citeauthoryear{Athreya et al.}{2002}]{Ath02} Athreya R. M., Mellier Y., van Waerbeke L., Pell\'{o} R., Fort B., Dantel-Fort M., 2002, A\&A, 384, 743
\bibitem[\protect\citeauthoryear{Bacon D.J. et al.}{2006}]{bgrt} Bacon D. J., Goldberg D. M., Rowe B. T. P., Taylor A. N., 2006, MNRAS, 365, 414
\bibitem[\protect\citeauthoryear{Carlberg et al.}{1997}]{Carl97} Carlberg R. G., et al., 1997, ApJL, 485, 13
\bibitem[\protect\citeauthoryear{Corless et al.}{2009}]{Corless09} Corless V. L., King L. J., Clowe D., 2009, MNRAS, 393, 1235
\bibitem[\protect\citeauthoryear{Czoske et al.}{2008}]{Cz08} Czoske O., Barnab\`{e} M., Koopmans L. V. E., Treu T., Bolton A. S., 2008, MNRAS, 384, 987
\bibitem[\protect\citeauthoryear{Dye et al.}{2008}]{Dye08} Dye S., Evans N. W., Belokurov V., Warren S. J., Hewett P., 2008, MNRAS, 388, 384
\bibitem[\protect\citeauthoryear{Gavazzi et al.}{2007}]{Gav07} Gavazzi R., Treu T., Rhodes J. D., Koopmans L. V. E., Bolton A. S., Burles S., Massey R. J., Moustakas L. A., 2007, ApJ, 667, 176
\bibitem[\protect\citeauthoryear{Goldberg D.M. and Bacon D.J.}{2005}]{gb05} Goldberg D. M., Bacon D. J., 2005, ApJ, 619, 741
\bibitem[\protect\citeauthoryear{Goldberg D.M. and Leonard A.}{2007}]{gl07} Goldberg D. M., Leonard A., 2007, ApJ, 660, 1003
\bibitem[\protect\citeauthoryear{Hansen et al.}{2005}]{Hansen05} Hansen S. M., McKay T. A., Wechsler R. H., Annis J., Sheldon E. S., Kimball A., 2005, ApJ, 633, 122
\bibitem[\protect\citeauthoryear{Hawken \& Bridle}{2009}]{Hawken09} Hawken A. J., Bridle S. L., 2009, MNRAS submitted, arXiv:0903.3938
\bibitem[\protect\citeauthoryear{Katgert et al.}{2004}]{Kat04} Katgert P., Biviano A., Mazure A., 2005, ApJ, 600, 657
\bibitem[\protect\citeauthoryear{Koopmans et al.}{2006}]{Koop06} Koopmans L. V. E., Treu T., Bolton A. S., Burles S., Moustakas L. A., 2006, ApJ, 649, 599
\bibitem[\protect\citeauthoryear{Lasky \& Fluke}{2009}]{LF09} Lasky P. D., Fluke C. J., 2009, MNRAS, 396, 2257
\bibitem[\protect\citeauthoryear{Leonard A. et al.}{2007}]{LGHM} Leonard A., Goldberg D. M., Haaga J. L., Massey R., 2007, ApJ, 666, 51
\bibitem[\protect\citeauthoryear{Leonard A. et al.}{2009}]{LKW09} Leonard A., King L. J., Wilkins S. M., 2009, MNRAS, 395, 1438
\bibitem[\protect\citeauthoryear{Lin et al.}{2004}]{Lin04} Lin Y.-T., Mohr J. J., Stanford S. A., 2004, ApJ, 610, 745
\bibitem[\protect\citeauthoryear{Lokas et al.}{2006}]{Lokas06} {\L}okas E. L., Prada F., Wojtak R., Moles M., Gottl\"{o}ber S., 2006, MNRAS, 366, L26
\bibitem[\protect\citeauthoryear{Mandelbaum et al.}{2008}]{Man08} Mandelbaum R., Seljak U., Hirata C. M., 2008, JCAPP, 8, 6
\bibitem[\protect\citeauthoryear{Navarro J.F. et al.}{1997}]{nfw} Navarro J. F., Frenk C. S., White S. D. M., 1997, ApJ, 490, 493
\bibitem[\protect\citeauthoryear{Newman et al.}{2009}]{Newman09} Newman A. B. et al., 2009, ApJ in press, arXiv:0909.3527
\bibitem[\protect\citeauthoryear{Okabe et al.}{2009}]{Okabe09} Okabe N., Takada M., Umetsu K., Futamase T., Smith G. P., PASJ submitted, arXiv:0903.1103
\bibitem[\protect\citeauthoryear{Okura Y. et al}{2007}]{okura1} Okura Y., Umetsu K., Futamase T., 2007, ApJ, 660, 995
\bibitem[\protect\citeauthoryear{Okura Y. et al}{2008}]{okura2} Okura Y., Umetsu K., Futamase T., 2008, ApJ, 680, 1
\bibitem[\protect\citeauthoryear{Rines \& Diaferio}{2006}]{RD06} Rines K., Diaferio A., 2006, AJ, 132, 1275
\bibitem[\protect\citeauthoryear{Rusin et al.}{2003}]{rusin03} Rusin D., Kochanek C. S., Keeton C. R., 2003, ApJ, 595, 29
\bibitem[\protect\citeauthoryear{Rusin \& Kochanek}{2005}]{RK05} Rusin D., Kochanek C. S., 2005, ApJ, 623, 666
\bibitem[\protect\citeauthoryear{Schneider P.}{1996}]{s96} Schneider P., 1996, MNRAS, 283, 837
\bibitem[\protect\citeauthoryear{Shaw et al.}{2006}]{Shaw06} Shaw L. D., Weller J., Ostriker J. P., Bode P., 2006, ApJ, 646, 815
\bibitem[\protect\citeauthoryear{Springel et al.}{2005}]{S05} Springel V., et al., 2005, Nature, 435, 629 
\bibitem[\protect\citeauthoryear{Treu \& Koopmans}{2002}]{TK02} Treu T., Koopmans T. V. E., 2002, ApJ, 575, 87
\bibitem[\protect\citeauthoryear{Tu et al.}{2009}]{Tu09} Tu H., et al., 2009, A\&A, 501, 475
\bibitem[\protect\citeauthoryear{van der Marel et al.}{2000}]{vdm00} van der Marel R. P., Magorrian J., Carlberg R. G., Yee H. K. C, Ellingson E., 2000, 119, 2038
\bibitem[\protect\citeauthoryear{Wang\&White}{2009}]{ww09}Wang J., White S. D. M., 2009, MNRAS, 396, 709
\bibitem[\protect\citeauthoryear{Wojtak et al.}{2007}]{Woj07} Wojtak R., {\L}okas E. L., Mamon G. A., Gottl\"{o}ber S., Prada F., Moles M., A\&A, 466, 437
\end{thebibliography}
\end{document}